\documentclass[a4paper,fleqn,usenatbib]{mnras}

\usepackage{mathptmx}

\usepackage[T1]{fontenc}
\usepackage{ae,aecompl}

\usepackage{graphicx}	
\usepackage{amsmath}	
\usepackage{amssymb}	

\newcommand\mP{\mathcal{P}}
\newcommand\mG{\mathcal{G}}
\newcommand\mB{\mathcal{B}}
\newcommand\mQ{\mathcal{Q}}
\newcommand\mM{\mathcal{M}}
\newcommand\mC{\mathcal{C}}
\newcommand\mA{\mathcal{A}}
\newcommand\mK{\mathcal{K}}
\newcommand\mE{\mathcal{E}}
\newcommand\mW{\mathcal{W}}
\newcommand\mS{\mathcal{S}}
\newcommand\mT{\mathcal{T}}
\newcommand\vpa{{v_\parallel}}
\newcommand\vx{{\bf x}}
\newcommand\vr{{\bf r}}

\title[Pairwise velocity model covering large and small scales]{Improving the modelling of redshift-space distortions - II.\\
A pairwise velocity model covering large and small scales} 

\author[D.  Bianchi, W. J. Percival \& J. Bel]{
Davide Bianchi,$^{1,3}$\thanks{E-mail: davide.bianchi@port.ac.uk}
Will J. Percival,$^{1}$
and Julien Bel$^{2,3}$
\\
$^{1}$Institute of Cosmology \& Gravitation, Dennis Sciama Building, University of Portsmouth, Portsmouth, PO1 3FX, UK\\
$^{2}$Aix Marseille Univ, Univ Toulon, CNRS, CPT, Marseille, France\\
$^{3}$INAF -- Osservatorio Astronomico di Brera, via Emilio Bianchi 46, I-23807 Merate, Italy
}

\date{Accepted XXX. Received YYY; in original form ZZZ}

\pubyear{2016}

\begin{document}
\label{firstpage}
\pagerange{\pageref{firstpage}--\pageref{lastpage}}
\maketitle

\begin{abstract}
We develop a model for the redshift-space correlation function, valid for both dark matter particles and halos on scales $>5\,h^{-1}$Mpc.
In its simplest formulation, the model requires the knowledge of the first three moments of the line-of-sight pairwise velocity distribution plus two well-defined dimensionless parameters.
The model is obtained by extending the Gaussian-Gaussianity prescription for the velocity distribution, developed in a previous paper, to a more general concept allowing for local skewness, which is required to match simulations.
We compare the model with the well known Gaussian streaming model and the more recent Edgeworth streaming model.
Using N-body simulations as a reference, we show that our model gives a precise description of the redshift-space clustering over a wider range of scales.
We do not discuss the theoretical prescription for the evaluation of the velocity moments, leaving this topic to further investigation.
\end{abstract}




\section{Introduction}

The large-scale structure (LSS) of the universe is the result of a
continuous infall process in which the peculiar velocity flows induced
by gravitational instability drive matter towards denser regions, thus amplifying primordial density fluctuations.
Peculiar velocities leave a characteristic imprint, known as ``redshift space distortions" \citep[RSD,][]{kaiser1987}, on the galaxy clustering pattern measured by redshift surveys \citep[see][for a review]{hamilton1998}.
If properly modelled, measurements of RSD provide a powerful way to constrain
fundamental cosmological parameters in the $\Lambda$CDM paradigm or
to search for evidences of deviations from this standard scenario.

The effects of RSD on the observed galaxy correlation function can be
summarised as follows.  On large scales the dominant contribution is
given by the coherent movement of galaxies towards overdense regions, such
as clusters, walls and filaments, and away from voids. This
``squashes'' the iso-correlation contours along the line of sight.  As
we move to smaller scales, the disordered motion of galaxies inside
those formed structures becomes increasingly important, resulting in
elongated iso-contours along the line of sight, usually referred to as
``fingers of God" \citep{jackson1972}.

Since the 1987 seminal work by Kaiser, significant efforts have been
made to model the redshift-space large-scale profile of the
correlation function and its Fourier counterpart, the power spectrum
\citep[e.g.][]{matsubara2008, taruya2010, reid2011, seljak2011,
  uhlemann2015}.  The standard approach is to use perturbation theory
(PT), to compute the density and velocity field to higher order
\citep[see e.g.][for a review]{bernardeau2002}.

Less explored is the small-scale behaviour of RSD where the density
contrast becomes comparable to unity, causing the breakdown of any
perturbative-expansion scheme.  As a way around this issue, a few
alternative approaches have been suggested, spanning
from analytic \citep[e.g.][]{sheth1996} to hybrid techniques in which
N-body simulations are used to tune fitting functions
\citep[e.g.][]{tinker2007, kwan2012} or as a reference realisation of
the redshift-space clustering, in which small departures from the
assumed $\Lambda$CDM cosmology can be mimicked by varying appropriate
halo-occupation-distribution (HOD) parameters \citep[][]{reid2014}.  A
good understanding of this small-scale limit is desirable for two main
reasons: (i)~It is rich in cosmological information, in particular if
our goal is to discriminate between different gravity model.
Specifically, it has been shown that modified gravity strongly affects
the pairwise velocity dispersion on these scales \citep{fontanot2013,
  hellwing2014}.  (ii)~The smaller the separation the higher the
signal-to-noise ratio and the less the cosmic variance, i.e. smaller
statistical error. Thus understanding this process allows us to push
measurements of the structure growth effects to smaller scales.

With this work we provide a framework in which these large- and
small-scale RSD processes can both be included, so that all
available information can be coherently extracted form redshift
surveys. We start from the ``streaming model" \citep{davis1983,
  fisher1995, scoccimarro2004b}, which describes how the redshift-space correlation function $\xi_S(s_\perp, s_\parallel)$ is modified with respect to its isotropic real-space counterpart $\xi_R(r)$: 
\begin{equation}
\label{eq streaming}
1 + \xi_S(s_\perp , s_\parallel) = \int dr_\parallel \ [1 + \xi_R(r)] \ \mP(r_\parallel - s_\parallel | \vr) \ ,
\end{equation}
where we have ignored wide-angle effects. Here $r^2 = {r_\perp}^2 + {r_\parallel}^2$ and $r_\perp=s_\perp$.
The effect of the velocity flows on the observed clustering is encoded in the pairwise line-of-sight velocity distribution function $\mP(v_\parallel |\vr) = \mP(r_\parallel - s_\parallel | \vr)$, which has a non trivial dependence on the separation ${\bf r}$.
Clearly, a proper modelling of this PDF is the key ingredient in the description.
Starting from this consideration, in
\citet{bianchi2015}, hereafter Paper I, we modelled the velocity PDF
by introducing the concept of Gaussian Gaussianty (GG), in which the
overall PDF is interpreted as a superposition of local Gaussian
distributions, whose mean and standard deviation are, in turn, jointly
distributed according to a bivariate Gaussian.  Here we extend that
line of research by introducing the more general concept of Gaussian
quasi-Gaussianity (GQG) and making explicit the dependence of the
velocity PDF on quantities that can be predicted by theory, namely its
first three moments.  We do not discuss which theoretical scheme
should be preferred for their evaluation, but rather we directly
measure these quantities from N-body simulations.  Our analysis
matches simulations over a large portion of the parameter space,
including redshifts from $z=0$ to $z=1$, dark matter (DM) particles,
halos with mass down to $10^{12}h^{-1}M_\odot$ and scales down to
$0h^{-1}$Mpc separation.  For all these configurations, we compare the
performance of our model with two different implementations of the
streaming model: the well known Gaussian streaming model (GSM), in
which a univariate-Gaussian profile is assumed for the velocity PDF
\citep{reid2011}; the more recent Edgeworth streaming model
\citep[ESM,][]{uhlemann2015}, in which the skewness is added to this
simple Gaussian picture by means of an Edgeworth expansion \citep[see
e.g.][]{blinnikov1998}.  We show that, under the GQG assumption, a
more precise description of the redshift-space clustering is obtained.

The paper is organised as follows.
In Sec. \ref{sec theory} we introduce our model.
As this work is the second in a series, the derivation we present follows the ``historical process'' that led us to introduce GQG.
In sections \ref{sec GG} - \ref{sec skew problem} we first review how to build a model based on GG, introduced in Paper I, and then show the (unexpected) limitations of such approach.
In the remainder of Sec. \ref{sec theory} we show how to overcome this issue.
Our final model is based on a few assumptions, which are referred as ansatze throughout the manuscript.
Given this, the model we are proposing should be considered a functional form for the velocity PDF that, irregardless of its derivation, incorporates all the fundamental features observed in both simulations and galaxy surveys, including exponential tails and skewness.
If required this GQG distribution can be exactly shaped into a Gaussian, which means that, by construction, the resulting streaming model is a generalisation of the widely used Gaussian streaming model.
Furthermore, we show that this PDF has the nontrivial property of being expressible as a functions of its first three moments, thus providing an explicit link to perturbation theory.
We believe that such a distribution would have been interesting to be studied even if it were unmotivated from a physical point of view, as sometimes happens in the literature.
This is of course not the case with GQG, which is explicitly derived based on considerations on how the overall PDF can be decomposed in local PDFs, with the spirit of keeping only the features of these latter that are relevant for RSD.
Using N-body simulations as a reference, in Sec. \ref{sec sim} we compare the performance of our model with that of GSM and ESM.
The primary purpose of this comparison is to show that, once the first three moments are given, the remaining degrees of freedom can be effectively absorbed in two numbers, the $\kappa$ parameters (see Sec. \ref{sec simple}).
Our results are summarised in Sec. \ref{sec discu}.
Details on how we measure physical quantities from the simulations are reported in the appendices.
Also in the appendices we discuss ideas for further developments.

\section{Modelling}\label{sec theory}

\subsection{GG distribution}\label{sec GG}

In paper I, we proposed a functional form for the line-of-sight pairwise velocity distribution,  
\begin{equation}
\label{eq gau1biv2}
 \mP(\vpa) = \int d\mu \ d\sigma \ \mG(\vpa|\mu,\sigma) \ \mB(\mu,\sigma) \ ,
\end{equation}
where
\begin{equation}
\label{eq gau1}
\mG(\vpa|\mu,\sigma) = \frac{1}{\sqrt{2\pi\sigma^2}}
 \exp\left[{-\frac{{(\vpa-\mu)}^2}{2 \sigma^2}}\right] \ ,
\end{equation} 
\begin{equation}
\mB(\mu,\sigma) = \dfrac{1}{2 \pi \sqrt{\det(C)}} \exp
\left[-\frac{1}{2} \Delta^T C^{-1} \Delta \right] \ ,
\end{equation}
\begin{equation}
 \Delta =
 \begin{pmatrix}
  \mu - M_\mu \\
  \sigma - M_\sigma
 \end{pmatrix}
 \qquad
 C =
 \begin{pmatrix}
  C_{\mu\mu} & C_{\mu\sigma} \\
  C_{\mu\sigma} & C_{\sigma\sigma} 
 \end{pmatrix} \ .
\end{equation}
The interpretation of Eq. (\ref{eq gau1biv2}) is straightforward: at any given separation $(r_\perp, r_\parallel)$, the overall velocity distribution can be approximated by a superposition of univariate Gaussians whose mean $\mu$ and standard deviation $\sigma$ are, in turn, jointly distributed as a bivariate Gaussian.
$M_\mu$ and $M_\sigma$ represent the mean of $\mu$ and $\sigma$, respectively, whereas $C$ is their covariance matrix. 
We showed in paper~I that the simple picture in which these univariate Gaussians represent local velocity distributions gives a good match to N-body simulations. 
Note that hereafter we write $\mP(\vpa)=\mP(\vpa|{\bf r})$, where the dependence on the separation is omitted for brevity, but still present in our model.
Specifically, it is encoded in how the parameters $M_\mu=M_\mu({\bf r})$, $M_\sigma = M_\sigma({\bf r})$ and $C=C({\bf r})$ vary with the separation.

\begin{figure}
 \begin{center}
   \includegraphics[width=9cm]{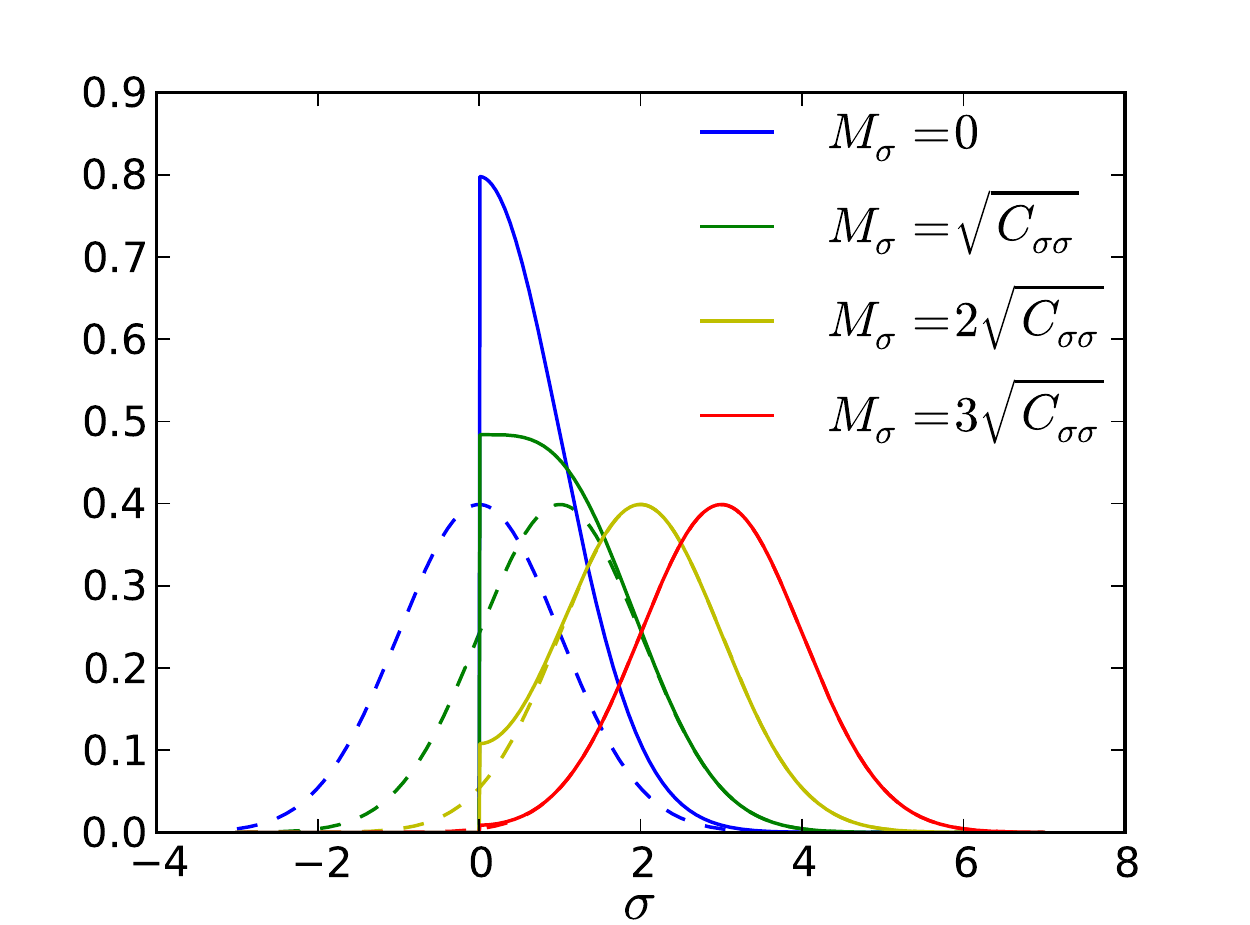}
    \caption{Cuts of $\mB$ (dashed) vs cuts of $\mB^\pm$ (solid) along the $\mu$ axis for different values of mean and variance (of $\sigma$), $M_\sigma$ and $C_{\sigma\sigma}$, respectively. For simplicity, we only report the case $M_\mu=C_{\mu\mu}=C_{\mu\sigma}=0$, but this behaviour is general.}
  \label{fig mirrorgau}
 \end{center}
\end{figure}

Strictly speaking, the above modelling is physically meaningful only if $M_\sigma \gtrsim 3 \sqrt{C_{\sigma\sigma}}$, i.e. only if the whole power of the bivariate Gaussian is limited to the positive $\sigma$ plane. 
To ensure that the expression is well behaved for $M_\sigma  \rightarrow 0$, we adopt for $\mG$ the normalisation factor $\sqrt{2\pi\sigma^2}$ rather than  $\sqrt{2\pi}\sigma$, Eq. (\ref{eq gau1}),
and we no longer have to deal with negative local distributions, independently of the width of the bivariate Gaussian $C_{\sigma\sigma}$.
We can write
\begin{align}
\label{eq gau1biv2gen}\mP(\vpa) &= \int^{+\infty}_{-\infty} d\mu \left(\int^0_{-\infty} d\sigma \ \mG \ \mB+ \int^{+\infty}_0 d\sigma \ \mG \ \mB \right) \nonumber \\
&= \int d\mu \ d\sigma \ \mG(\vpa|\mu,\sigma)  \ \mB^\pm(\mu,\sigma) \ ,
\end{align}
where we have defined 
\begin{equation}\label{eq bpm}
\mB^\pm(\mu,\sigma) \equiv \begin{cases}
\mB(\mu,-\sigma) + \mB(\mu,\sigma) & \sigma \geq 0\\
0 & \sigma < 0 
\end{cases} \ .
\end{equation}
Eq. (\ref{eq gau1biv2gen}) generalises Eq. (\ref{eq gau1biv2}) in a natural way, such that all the fundamental properties of $\mP$ are conserved. In particular the relation between moments of $\mP$ and  $\mB$, presented in the first column of Tab.~\ref{tab mvsM}, remains valid in this more general formulation.
Still, it is appropriate to note that the moments of $\mB^\pm$ differ from those of $\mB$, and they coincide in the limiting case in which $M_\sigma \gtrsim 3 \sqrt{C_{\sigma\sigma}}$ \footnote{
Following the notation introduced in Tab. \ref{tab momdef}, when $M_\sigma \lesssim 3 \sqrt{C_{\sigma\sigma}}$, for the first non-central moments it holds that
\begin{align}
M^{(1)}_0 &= M_\mu \qquad M^{(1)}_1 \ne M_\sigma \nonumber \ ,
\end{align}
whilst for the second central moments,
\begin{align}
C^{(2)}_{00} &= C_{\mu\mu} \qquad C^{(2)}_{11} \ne C_{\sigma\sigma} \qquad C^{(2)}_{01} \ne C_{\mu\sigma} \nonumber \ .
\end{align}
More in general, $\mB$ and $\mB^\pm$ share by construction all the even non-central moments $M^{(2n)}_{k_1 \cdots k_{2n}}$.
}.
See Tab. \ref{tab momdef} for the definitions of the moments.
\begin{table*}
 \centering
  \begin{tabular}{cll}
  &&\\
   \multicolumn{1}{c}{\Large PDF} & \multicolumn{1}{c}{\Large moments} & \multicolumn{1}{c}{\Large central moments} \\
  \hline
  \hline
  &&\\
  $\mP$ & $m^{(n)} \equiv \int d\vpa \ \vpa^n \ \mP(\vpa)$ & $c^{(n)} \equiv \int d\vpa \ \left[\vpa-m^{(1)} \right]^n \ \mP(\vpa)$ \\
  &&\\
   $\mB$ & $M_\mu \equiv \int d\mu d\sigma \ \mu \ \mB(\mu,\sigma)$ & $C_{\mu\mu} \equiv \int d\mu d\sigma \ {(\mu-M_\mu)}^2 \ \mB(\mu,\sigma)$ \\
  &&\\
    & $M_\sigma \equiv \int d\mu d\sigma \ \sigma \ \mB(\mu,\sigma)$ & $C_{\sigma\sigma} \equiv \int d\mu d\sigma \ {(\sigma-M_\sigma)}^2 \ \mB(\mu,\sigma)$ \\ 
  &&\\
    & &  $C_{\mu\sigma} \equiv \int d\mu d\sigma \ (\mu-M_\mu) (\sigma-M_\sigma) \ \mB(\mu,\sigma)$ \\
  &&\\
   $\mB^\pm$ & $M_{k_1,\cdots,k_n}^{(n)} \equiv \int d\mu d\sigma \ \mu^{n-\sum_i k_i} \sigma^{\sum_i k_i} \ \mB(\mu,\sigma)$ & $C_{k_1,\cdots,k_n}^{(n)} \equiv \int d\mu d\sigma \ {\left[\mu-M^{(1)}_0\right]}^{n-\sum_i k_i} {\left[\sigma-M^{(1)}_1\right]}^{\sum_i k_i} \ \mB^\pm(\mu,\sigma)$ \\
  &&\\
 \end{tabular}
 \caption{Definitions and notation adopted to describe the moments of the probability distribution functions considered in this work: $\mP$, $\mB$ and $\mB^\pm$. 
   $n$ is the order of the moment and $k \in \{0,1\}$. Not to overcomplicate the notation, for $\mB$ we only define the first non-central and second central moments.}
 \label{tab momdef}
\end{table*}
In Fig. \ref{fig mirrorgau} we show the comparison between $\mB$ and $\mB^\pm$ for a few selected cuts in $\mu$.

Although the physical meaning of the GG distribution is well described by  Eq. (\ref{eq gau1biv2gen}), it is important to note that $\mu$ can be integrated analytically. The integration gives 
\begin{align}
\label{eq GG1dim}
\mP(\vpa) = \int d \sigma \frac{1}{2 \pi \mA} \exp \left[-\frac{\sum_{n=0}^2 \mK_n \ {(\vpa - M_\mu)}^n }{2 \mA^2} \right] \ ,
\end{align}
where
\begin{align}
\mA^2 &= C_{\sigma\sigma} \sigma^2 + C_{\mu\mu} C_{\sigma\sigma} - {C_{\mu\sigma}}^2 \\
\mK_2 &=  C_{\sigma\sigma} \\
\mK_1 &= - 2 C_{\mu\sigma} \left(\sigma - M_\sigma \right) \\
\mK_0 &=  \left(\sigma^2 + C_{\mu\mu}\right) {\left(\sigma - M_\sigma \right)}^2 \ .
\end{align}
This result is particularly useful from a numerical point of view since, for any given set of parameters $\{M_\mu, M_\sigma, C_{\mu\mu}, C_{\sigma\sigma}, C_{\mu\sigma}\}$, it allows us to compute $\mP$ via a simple, i.e. fast, 1-dimensional integration.

Following standard practice, we define the moment generating function (MGF) as
\begin{equation}
\mM(t) = \langle e^{t \vpa} \rangle = \int d \vpa \ e^{t \vpa} \ \mP(\vpa) \ .
\end{equation}
One important property of the MGF is that it allows us to compute the moments iteratively at any order,
\begin{equation}
m^{(n)} = \frac{d^n \mM}{d t^n} \bigg|_{t=0} \ .
\end{equation}
For the GG distribution we get
\begin{equation}
\mM(t) = \frac{1}{\sqrt{1-t^2 C_{\sigma\sigma}}} \exp \left[t M_\mu + \frac{1}{2} t^2 {M_\sigma}^2 + \Theta(t) \right] \ ,
\end{equation}
where
\begin{align}
\Theta(t) &= \frac{1/2}{1-t^2 C_{\sigma\sigma}} \bigg[t^2 C_{\mu\mu} + 2 t^3 C_{\mu\sigma} M_\sigma \nonumber \\
&\qquad \qquad + t^4 \left({M_\sigma}^2 C_{\sigma\sigma} - \det{C}\right)\bigg] \ .
\end{align}
Similarly, we can define define the cumulant generating function, $\mC(t) = \log \langle e^{t \vpa} \rangle$, which, for the GG distribution, takes the form
\begin{equation}
\mC(t) = t M_\mu + \frac{1}{2} t^2 {M_\sigma}^2 + \Theta(t) - \frac{1}{2}\ln\left(1 - t^2 C_{\sigma\sigma} \right) \  .
\end{equation} 
In the following we briefly discuss a few cases of interest corresponding to particular combinations of the parameters of the bivariate Gaussian. 
\begin{enumerate}
\item
If $C_{\sigma\sigma}  = C_{\mu\sigma} = 0$ we get
\begin{equation}
\mM(t) = \exp \left[t M_\mu + \frac{1}{2} t^2 \left({M_\sigma}^2 + C_{\mu\mu} \right)\right]\ ,
\end{equation}
which is the MGF of a Gaussian with mean $M_\mu$ and variance ${M_\sigma}^2 + C_{\mu\mu}$.
In other words, the superposition of fixed-variance Gaussians (i.e. $C_{\sigma\sigma}  = 0$) is, in turn, a Gaussian.
We now consider the two limiting cases, $C_{\mu\mu} = 0$ and $M_\sigma = 0$.
From a physical point of view, $C_{\mu\mu} = 0$ corresponds to a scenario in which at any position in the Universe we measure the same pairwise velocity PDF, which is clearly what we expect in the large scale limit\footnote{In this description the PDF is Gaussian because we are using local Gaussians as building blocks (in the following this assumption will be slightly relaxed). As discussed in Sec. \ref{sec simple}, even on large scales the true velocity PDF is never exactly Gaussian. Nonetheless, the Gaussian approximation becomes more and more accurate as the separation increases. In practice we will use this limit as an ''infinite-scale'' limit, which is never really reached.}.
On the other hand, the limit $M_\sigma = 0$ represents a superposition of Dirac deltas whose mean is Gaussian distributed.  
Such a scenario is not compatible with any reasonable pairwise velocity PDF, although it might be useful for different applications, e.g. when describing the time evolution of the 1-particle velocity PDF.
More explicitly, in the phase-space formalism, it is commonly assumed that, at any position, the 1-particle velocity density (more precisely the momentum density) is well approximated by a Dirac delta (the so called single-flow approximation).
After shell crossing this assumption is no longer valid and we have to resort to distributions with a broader profile, e.g. Gaussians.
Whether the evolution of these distributions can be captured by a bivariate Gaussian description of their mean and variance, or, in other words whether the statistics of a fluid can be described by a GG distribution is an interesting question that we will try to answer in a further work.
\item
In the very-small-scale limit the statistics are dominated by virialized regions, which implies negligible local infall velocity, i.e. $M_\mu = C_{\mu\mu} = C_{\mu\sigma} = 0$.
The corresponding MGF is
\begin{equation}
\label{eq smallscale}
\mM(t) = \frac{1}{\sqrt{1-t^2 C_{\sigma\sigma}}} \exp \left(\frac{ \frac{1}{2} t^2 {M_\sigma}^2 }{1-t^2 C_{\sigma\sigma}} \right) \  .
\end{equation}
As shown in appendix \ref{app momgenfun}, when ${M_\sigma}^2 = 2C_{\sigma\sigma}$ this latter approximate the MGF of an exponential distribution.
It is well know from simulations and observations \citep[e.g.][]{zurek1994,davis1983} that the small-scale velocity PDF is nearly exponential and is therefore important that this limit is included in our description, although we will not explicitly use it in our modelling (but see App. \ref{sec smallscale}).     
\end{enumerate}

Finally, it is worth mentioning two potentially relevant applications of the GG-distribution MGF:  
\begin{enumerate}
\item
It can be used to compute the velocity PDF via a simple fast Fourier transform, which is computationally attractive. 
\item
It allows us to directly model the redshift-space power spectrum, see e.g. Eq. (13) in \citet{scoccimarro2004b}.
\end{enumerate}

\subsection{Strategy}\label{sec strategy}

If we assume that the true velocity PDF is well approximated by the GG distribution, Eq. (\ref{eq gau1biv2}), or equivalently Eq. (\ref{eq gau1biv2gen}), we can think of using this model to extract cosmological information from galaxy redshift surveys via RSD.  

Since the five scale-dependent parameters $M_\mu$, $M_\sigma$, $C_{\mu\mu}$, $C_{\sigma\sigma}$ and $C_{\mu\sigma}$ on which the distribution depends have a clear interpretation, we can think of directly predicting them. 
One intriguing aspect of such an approach is that it allows us to reason in terms of local distributions, suggesting the possibility of naturally including a multi-stream description.
In general, such an issue is expected to become more and more important as we want to describe the small-scale nonlinear regime.
Roughly speaking, an extension form single- to multi-flow scenario could be obtained by using Gaussians instead of Dirac delta distributions for the local 1-particle velocity PDF (more properly for the momentum part of the phase-space distribution function).
The resulting local pairwise velocity distribution will then be Gaussian as well and, as a consequence, the overall pairwise velocity PDF will be compatible with the GG prescription.   
We leave these considerations to further work.

Instead, we follow the conceptual spirt of the Gaussian streaming model, as implemented by \citet{reid2011}. 
The Gaussian streaming model (GSM) relies on the assumption that, at a any given separation $(r_\perp, r_\parallel)$, the overall line-of-sight pairwise velocity PDF is well approximated by an univariate Gaussian, whose mean and variance were obtained by \citet{reid2011} via PT.
This approach can be extended to include more general and realistic distributions, with more than two free moments.  
The $n$-th moment of the line-of sight pairwise velocity distribution $\mP(\vpa)$ is
\begin{equation}
m^{(n)} = \frac{\left\langle (1 + \delta_1) (1 + \delta_2) \ \vpa^n \right\rangle}{\left\langle (1 + \delta_1) (1 + \delta_2) \right\rangle} \ ,
\end{equation}
where $\delta_i = \delta(\vx_i)$, with $\delta$ being the usual density contrast.
Similarly, the central moments are defined as 
\begin{equation}
c^{(n)} = \frac{\left\langle (1 + \delta_1) (1 + \delta_2) {\left(\vpa- m^{(1)}\right)}^n \right\rangle}{\left\langle (1 + \delta_1) (1 + \delta_2) \right\rangle} \ .
\end{equation}
In principle, these quantities can be predicted by PT even for $n>2$ \citep[e.g.][see also App. \ref{sec smallscale} for a simple example of how these moments can be predicted on nonlinear scales]{juszkiewicz1998, uhlemann2015}.
The GG distribution includes the Gaussian distribution as a limiting case. 

By inverting the system in the lefthand column of Tab.~\ref{tab mvsM} we can write the bivariate Gaussian $\mB$ as a function of the first five moments of $\mP(\vpa)$, namely $m^{(1)}$, $c^{(2)}$, $c^{(3)}$, $c^{(4)}$ and $c^{(5)}$.
Explicit expressions for the resulting $M_\mu$, $M_\sigma$, $C_{\mu\mu}$, $C_{\sigma\sigma}$ and $C_{\mu\sigma}$ are reported in the righthand column of Tab.~\ref{tab mvsM}. 
\begin{table*}
 \centering
 \begin{tabular}{ll}
 $\mP$ vs. $\mB$ & $\mB$ vs. $\mP$ \\[2.5mm]
 \hline
 \hline \\[2.5mm]
  $m^{(1)} = M_\mu$ & $M_\mu =  m^{(1)}$  \\[5mm]
    $c^{(2)} = {M_\sigma}^2 + C_{\mu\mu} + C_{\sigma\sigma}$ & $C_{\sigma\sigma} = \frac{1}{20} \frac{c^{(5)}}{c^{(3)}} - \frac{1}{2} c^{(2)}$ \\[5mm]
  $c^{(3)} = 6 M_\sigma C_{\mu\sigma}$ & ${M_\sigma}^2 = \frac{c^{(4)} - 3 {c^{(2)}}^2 - 6 {C_{\sigma\sigma}}^2 + \sqrt{{\left[c^{(4)} - 3 {c^{(2)}}^2 - 6 {C_{\sigma\sigma}}^2 \right]}^2 - 16 C_{\sigma\sigma} {c^{(3)}}^2}}{24 C_{\sigma\sigma}}$ \\[5mm]
    $c^{(4)} = 3 \left({M_\sigma}^2 + C_{\mu\mu} \right)^2$ & $C_{\mu\mu} =  c^{(2)} - {M_\sigma}^2 - C_{\sigma\sigma}$ \\[2.5mm]
        $\qquad + 6 \left[C_{\sigma\sigma} \left(3{M_\sigma}^2 + C_{\mu\mu}\right)+ 2{C_{\mu\sigma}}^2\right] + 9 {C_{\sigma\sigma}}^2$ & \\[5mm]    
      $c^{(5)} = 60 M_\sigma C_{\mu\sigma} \left({M_\sigma}^2 + C_{\mu\mu} + 3C_{\sigma\sigma}\right)$ &  $C_{\mu\sigma} =  \frac{1}{6} \frac{c^{(3)}}{M_\sigma}$ \\[5mm]
       \end{tabular}
 \caption{Moments of the line-of-sight velocity distribution $\mP$ as a function of the moments of the bivariate Gaussian $\mB$ (left column) and viceversa (right column).} 
 \label{tab mvsM}
\end{table*}
The inversion is well defined as long as $c^{(3)} \neq 0$.
Formally, if $c^{(3)} = 0$ (which implies that all the odd central moments disappear as well), in order to have a one-to-one correspondence between $\mP$ and $\mB$ we need to include the 6th moment in the analysis.
We will see that this is not relevant in our modelling.

Although in principle the first five moments can be obtained via PT, in practice the complexity of the calculations grows rapidly with the order of both moment and perturbative expansion.
We therefore decide to adopt an hybrid approach in which we assume that the first three moments can be directly predicted, whilst 4th and 5th moment (and in general all higher order moments) are implicitly modelled as functions of a set of physically-meaningful dimensionless parameters, which arise naturally by general considerations about the properties of the GG distribution itself. The amplitude of these parameters is then obtained by comparisons with N-body simulations.

\subsection{Parameterisation}\label{sec model}

The expression for the first three moments of $\mP$ as a function of the moments of $\mB$ is
\begin{align}
m^{(1)} &= M_\mu \label{eq m1} \\
c^{(2)} &= {M_\sigma}^2 + C_{\mu\mu} + C_{\sigma\sigma} \label{eq c2} \\
c^{(3)} &= 6 M_\sigma C_{\mu\sigma} \label{eq c3} \ .
\end{align}
It is clear that the GG parameters $\{M_\mu, M_\sigma, C_{\mu\mu}, C_{\sigma\sigma}, C_{\mu\sigma}\}$ are uniquely defined once we specify $\{m^{(1)}, c^{(2)}, c^{(3)}\}$ plus a prescription on how to split $c^{(2)}$ into the three summands ${M_\sigma}^2$, $C_{\mu\mu}$ and $C_{\sigma\sigma}$, i.e. a prescription for their relative weight.
We then rewrite Eq. (\ref{eq c2}) in terms of three dimensionless quantities,
\begin{align}
 \label{eq fimp}
1 =  \frac{{M_\sigma}^2 }{c^{(2)} }+ \frac{C_{\mu\mu}}{c^{(2)} } + \frac{C_{\sigma\sigma}}{c^{(2)} } = \varphi_{M\sigma} + \varphi_{C\mu\mu} + \varphi_{C\sigma\sigma} \ .
\end{align}
Due to isotropy, $\mP(v_\parallel | r_\perp, r_\parallel)$ can be seen as the projection of a 2-dimensional distribution $\mP_r(v_r, v_t | r)$, where the subscripts $r$ and $t$ stand for parallel and perpendicular to the pair separation, see App.~\ref{app moments}.
As a consequence $c^{(2)}$ is in general characterised by the following symmetry,
\begin{equation}
\label{eq c2symm}
c^{(2)}(r,\mu_{\theta})=c^{(2)}_r(r) \ \mu_\theta^2 + c^{(2)}_t(r) \ \left(1 - \mu_\theta^2 \right) \ .
\end{equation}  
It is then convenient to define
\begin{align}
1 =& \ \varphi_{M\sigma}^{(r)}(r) + \varphi_{C\mu\mu}^{(r)}(r) + \varphi_{C\sigma\sigma}^{(r)}(r) \\
1 =& \ \varphi_{M\sigma}^{(t)}(r) + \varphi_{C\mu\mu}^{(t)}(r) + \varphi_{C\sigma\sigma}^{(t)}(r)
\end{align}
so that instead of three 2-dimensional functions we have to deal with six 1-dimensional functions\footnote{
This decomposition is based on the implicit assumption that the symmetry described by Eq. \ref{eq c2symm} can be applied not only to $c^{(2)}$ but also individually to each of its three building blocks,  i.e. $C_{\sigma\sigma}(r,\mu_\theta) = C_{\sigma\sigma}^{(r)}(r) \mu_\theta^2 + C_{\sigma\sigma}^{(t)}(r) \left(1-\mu_\theta^2\right)$, and simliarly for ${M_\sigma}^2$ and $C_{\mu\mu}$.
The $\varphi$ functions can be explicitly defined as $\varphi_{C\sigma\sigma}^{(r)} = C_{\sigma\sigma}^{(r)}/c^{(2)}_r$ and $\varphi_{C\sigma\sigma}^{(t)} = C_{\sigma\sigma}^{(t)}/c^{(2)}_t$.
It follows that $\varphi_{C\sigma\sigma} = \varphi_{C\sigma\sigma}^{(r)} \mu_\theta^2 + \varphi_{C\sigma\sigma}^{(t)} \left(1-\mu_\theta^2\right)$.
}.

Clearly, given the above equations, the functions we need to model are actually only four. 
A simple ansatz is then
\begin{align}
\varphi_{C\mu\mu}^{(r)}(r) &= \kappa_\mu^{(r)} \ g(r/r_g) \label{eq phi1}\\
\varphi_{C\sigma\sigma}^{(r)}(r) &= \kappa_\sigma^{(r)} \ g(r/r_g) \label{eq phi2}\\ 
\varphi_{M\sigma}^{(r)}(r) &= 1 - \varphi_{C\mu\mu}^{(r)}(r) - \varphi_{C\sigma\sigma}^{(r)}(r) \label{eq phi3}\\
\varphi_{C\mu\mu}^{(t)}(r) &= \kappa_\mu^{(t)} \ g(r/r_g) \label{eq phi4}\\
\varphi_{C\sigma\sigma}^{(t)}(r) &= \kappa_\sigma^{(t)} \ g(r/r_g) \label{eq phi5}\\ 
\varphi_{M\sigma}^{(t)}(r) &= 1 - \varphi_{C\mu\mu}^{(t)}(r) - \varphi_{C\sigma\sigma}^{(t)}(r) \label{eq phi6}\ ,
\end{align} 
where $g$ can be any monotonic regular function such that 
\begin{equation}
g(r) \rightarrow
\begin{cases}
0 & r \rightarrow \infty \\
1 & r\rightarrow 0 \\
\end{cases} \ ,
\end{equation}
e.g. $g(r/r_g)=\frac{1}{1+(r/r_g)^2}$.
By construction $r_g$ represents the scale above which the Gaussian limit is recovered, whereas $\kappa_\mu^{(r)}$, $\kappa_\sigma^{(r)}$, $\kappa_\mu^{(t)}$, $\kappa_\sigma^{(t)}$ represent the amplitudes of the corresponding $\varphi$ functions at $r=0$.

\subsection{The skewness problem}\label{sec skew problem}

\begin{figure}
 \begin{center}
   \includegraphics[width=9cm]{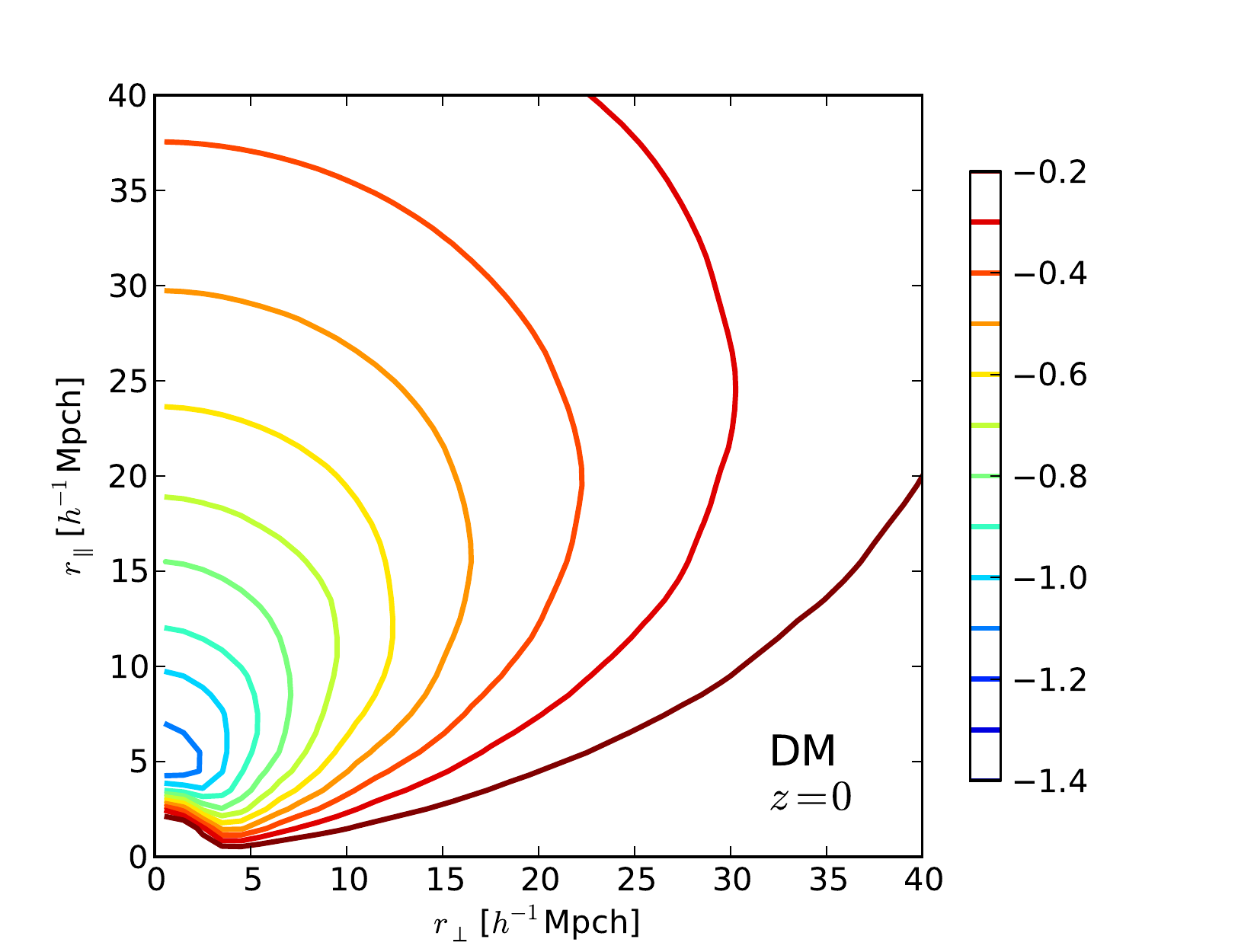}
   \includegraphics[width=9cm]{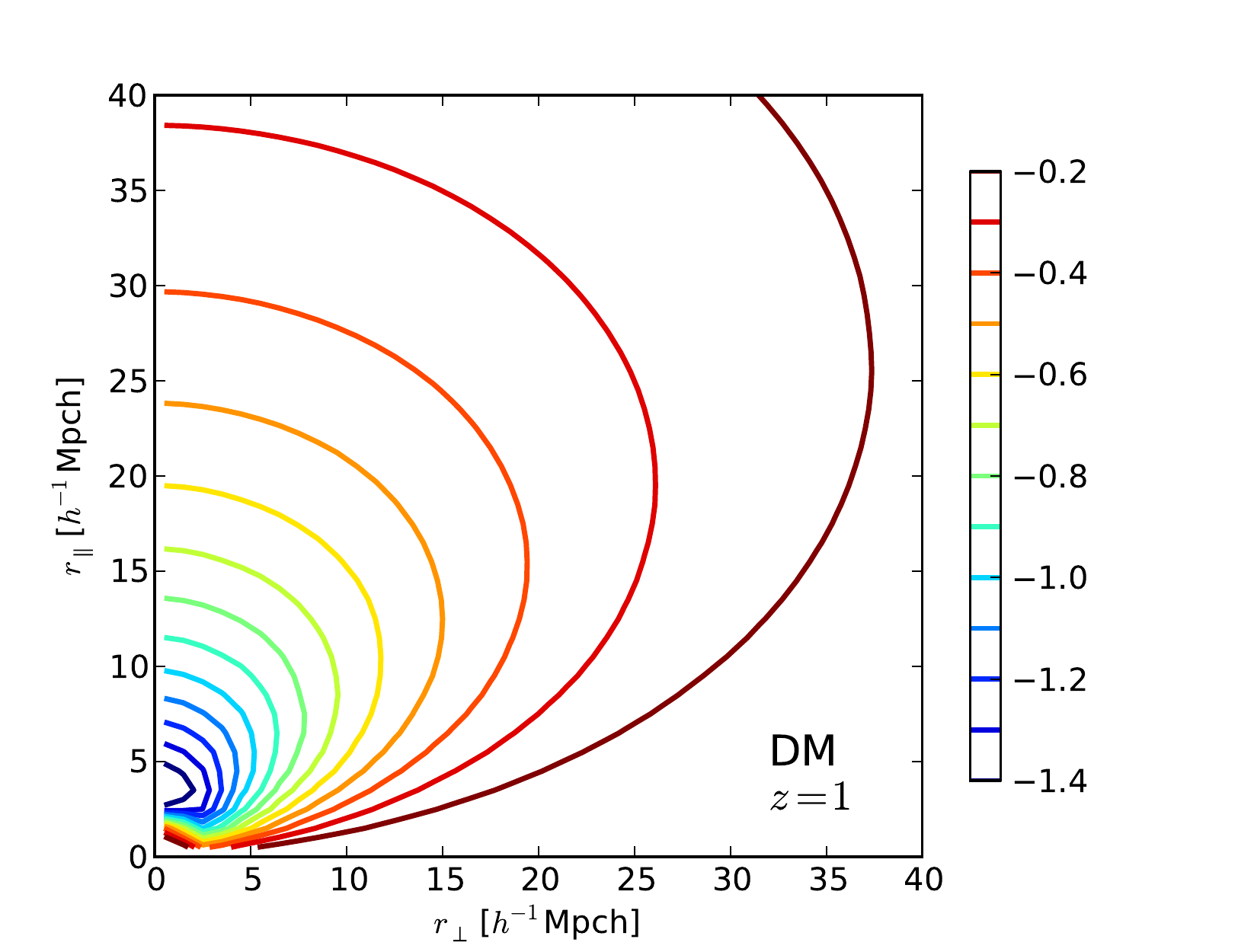}
   \caption{Measurement from simulations of the skewness of the
      line-of-sight pairwise velocity distribution, $\gamma =
      c^{(3)}/{c^{(2)}}^{3/2}$, presented using iso-skewness contours
      as a function of the real-space separation parallel and perpendicular to the line of sight, $r_\parallel$ and $r_\perp$, respectively, for DM particles from the MDR1 simulation at two different redshifts, as labelled in the figure.}
  \label{fig skewness}
 \end{center}
\end{figure}

Independently of the functional form chosen for $\varphi_{M\sigma}$, $\varphi_{C\mu\mu}$ and $\varphi_{C\sigma\sigma}$, from Eqs. (\ref{eq c3}) and (\ref{eq fimp}) we can write
\begin{equation}
\label{eq c3vsf}
c^{(3)} = 6 {c^{(2)}}^{3/2} \rho \ \sqrt{\varphi_{M\sigma} \ \varphi_{C\mu\mu} \ \varphi_{C\sigma\sigma}} \ ,
\end{equation}
where $\rho \equiv C_{\mu\sigma} / \sqrt{C_{\mu\mu} \ C_{\sigma\sigma}}$ is the correlation coefficient of the bivariate Gaussian.
Since in general $|\rho| < 1$, for any given $c^{(2)}$ Eq.~(\ref{eq c3vsf}) provides us with an upper bound for $|c^{(3)}|$,
\begin{equation}\label{eq skewupperlimit}
\left|c^{(3)}\right| < \frac{2}{\sqrt{3}}{c^{(2)}}^\frac{3}{2} \ , 
\end{equation}
corresponding to $\rho = \pm 1$ and $\varphi_{M\sigma} = \varphi_{C\mu\mu} = \varphi_{C\sigma\sigma} = 1/3$.
By explicitly defining the skewness, $\gamma \equiv c^{(3)}/{c^{(2)}}^{3/2}$, we have $\left| \gamma \right| < 2/ \sqrt{3} \sim 1.155$.
In Fig. \ref{fig skewness} we show that this limit is reached for dark matter at $z=0$ at $r \sim 5 h^{-1}$Mpc, $\mu_\theta \sim 0$, and is exceeded at higher redshift.
For halo catalogues, not shown in the figure for simplicity, this behaviour is even more marked\footnote{In Paper I this issue did not arise because only DM particles at $z=0$ were considered.}. 
Thus we see that the GG model is unable to match the observed level of skewness and requires further generalisation as described in the next section (but see also App. \ref{app alternative} for an alternative approach).

\subsection{Gaussian (local) quasi-Gaussianity}\label{sec GQG}

To overcome this problem we generalize GG by introducing the concept of Gaussian (local) quasi-Gaussianity (GQG).
We can account for a small deviation from local Gaussianity by Edgeworth-expanding the local distributions,
\begin{equation}
\label{eq GQG}
 \mP(\vpa) = \int d\mu \ d\sigma \ \mE(\vpa|\mu,\sigma) \ \mB(\mu,\sigma) \ ,
\end{equation}
where
\begin{align}
\label{eq edwo1}
\mE(\vpa|\mu,\sigma) =& \mG(\vpa|\mu,\sigma) \left[1 + \frac{\gamma_L}{6} \ H_3\left(\frac{\vpa - \mu}{\sigma} \right) \right]\ ,
\end{align}
\begin{equation}
\gamma_L = \frac{c^{(3)}_L}{\sigma^3} \ ,
\end{equation}
\begin{equation}\label{eq Hermite}
H_3(x) = x^3-3x \ .
\end{equation}
$c^{(3)}_L$ is third central moment of the local distribution $\mE$ (hence $\gamma_L$ is the local skewness) and $H_3$ is the third probabilistic Hermite polynomials.
It should be noted that, since the integral in Eq. (\ref{eq GQG}) formally includes negative values of $\sigma$, from Eqs. (\ref{eq edwo1}) and (\ref{eq Hermite}) it follows that both positively and negatively skewed (quasi) Gaussians contribute to the overall PDF.
This guaranties that the net contribution of the local to the overall skewness vanishes for $\sigma \rightarrow 0$, a desirable property if we want to avoid the nonsense of skewed Dirac deltas.   
It is nonetheless useful to say that, although the formulation of GQG clearly follows from the idea of allowing for a small skewness correction on local distributions, in a more general picture it can also be seen just as a generalised Edgeworth expansion, i.e. a practical way to control the skewness of a distribution without changing its first two moments.
In the perspective of using the model for a Montecarlo estimation of cosmological parameters in which second and third moments are free to vary, it is important to have removed a potential source of artefacts such those that would arise from exceeding the upper limit of Eq. (\ref{eq skewupperlimit}).     
As in the simpler case of the GG distribution, it is possible to integrate Eq.~(\ref{eq GQG}) with respect to $\mu$, 
\begin{align}
\label{eq GQG1dim}
\mP(\vpa) =& \ \int d \sigma \left\{\frac{1}{2 \pi \mA} \exp \left[-\frac{\sum_{n=0}^2 \mK_n \ {(\vpa - M_\mu)}^n }{2 \mA^2} \right] \nonumber \right.\\ 
&\left. \times \left[1+ \frac{\mS}{\mA^6}\sum_{k=0}^3 \mQ_k \ {(\vpa - M_\mu)}^k \right] \right\}\ ,
\end{align}
where $\mA$ and $\mK_n$ are defined in section \ref{sec GG}, and 
\begin{align}
\mS =& \ \gamma_L \ \sigma^3 / 6 \\
\mQ_3 =& \ {C_{\sigma\sigma}}^3\\
\mQ_2 =& \ - 3 C_{\mu\sigma} {C_{\sigma\sigma}}^2 \ (\sigma-M_\sigma)\\
\mQ_1 =& \ 3 C_{\sigma\sigma} [M_\sigma {C_{\mu\sigma}}^2 (M_\sigma - 2\sigma) \nonumber \\
&+ {C_{\mu\sigma}}^2 (\sigma^2 + C_{\sigma\sigma}) - {C_{\sigma\sigma}}^2 (\sigma^2 + C_{\mu\mu})] \\
\mQ_0 =& \ {C_{\mu\sigma}}^3 [3C_{\sigma\sigma}(M_\sigma -\sigma) - {(\sigma-M_\sigma)}^3] \nonumber \\
&+ 3C_{\mu\sigma} {C_{\sigma\sigma}}^2 (\sigma^2+C_{\mu\mu})(\sigma-M_\sigma) \ .
\end{align}
The first three moments of the GQG distribution are
\begin{align}
m^{(1)} =& \ M_\mu \\
c^{(2)} =& \ {M_\sigma}^2 + C_{\mu\mu} + C_{\sigma\sigma} \\
c^{(3)} =& \ 6M_\sigma C_{\mu\sigma} + \gamma_L M_\sigma ({M_\sigma}^2 + 3C_{\sigma\sigma})\label{eq skew GQG} \ .
\end{align}
These are the same as the GG distribution apart for the $\gamma_L M_\sigma ({M_\sigma}^2 + 3C_{\sigma\sigma})$  term which accounts for the excess skewness.
Keeping in mind that $c^{(3)}$ is given and that Eq. (\ref{eq skew GQG}) can be written as $c^{(3)} = 6\rho\sqrt{{M_\sigma}^2 C_{\mu\mu} C_{\sigma\sigma}} + \gamma_L M_\sigma ({M_\sigma}^2 + 3C_{\sigma\sigma})$, there are (at least) two practical ways to use the GQG prescription.
\begin{enumerate}
\item
We can define
\begin{equation}
\rho_0=\frac{c^{(3)}}{6\sqrt{{M_\sigma}^2 C_{\mu\mu} C_{\sigma\sigma}}} \ ,
\end{equation}
and adopt the following prescription,
\begin{align}
\begin{bmatrix}
\rho \\[2mm]
\gamma_L
\end{bmatrix}
 &=
\begin{bmatrix}
\rho_0 \\[2mm]
0
\end{bmatrix} \ ,
\end{align}
if $|\rho_0|<1$, whilst
\begin{align}\label{eq correction1}
\begin{bmatrix}
\rho \\[2mm]
\gamma_L
\end{bmatrix}
 &=
\begin{bmatrix}
\rho_0/|\rho_0| \\[2mm]
\frac{c^{(3)} - \rho_0/|\rho_0| \ 6 \sqrt{{M_\sigma}^2 C_{\mu\mu} C_{\sigma\sigma}}}{M_\sigma({M_\sigma}^2 + 3C_{\sigma\sigma})}
\end{bmatrix} \ ,
\end{align}
elswhere.
This corresponds to using GQG as an empirical correction for GG, to be "switched on" only when required by the third moment.
The benefit of this approach is that it does not require any additional parameter\footnote{Formally for $|\rho|=1$ a bivariate Gaussian is not well defined, therefore for any practical application we have to modify Eq. (\ref{eq correction1}) with $\rho=\frac{\rho_0}{|\rho_0|}-\epsilon$, where $\epsilon \ll 1$.}, the downside is that is not guaranteed that the shape of the velocity PDF varies smoothly with~$(r_\perp, r_\parallel)$.  
\item
The alternative is to use
\begin{align}\label{eq skew frac}
\begin{bmatrix}
\rho \\[4mm]
\gamma_L
\end{bmatrix}
 &=
\begin{bmatrix}
\alpha \ \frac{c^{(3)}}{6\sqrt{{M_\sigma}^2 C_{\mu\mu} C_{\sigma\sigma}}}  \\[4mm]
(1-\alpha) \ \frac{c^{(3)}}{M_\sigma ({M_\sigma}^2 + 3C_{\sigma\sigma})}
\end{bmatrix} \ ,
\end{align}
where, by construction, $\alpha \in (0,1)$ controls the ratio between the skewness created by the covariance $C_{\mu\sigma}$ and the local skewness.
In practice, rather then $\alpha$ we prefer to use the parameter $\tau \in (0,+\infty)$, defined as follows, 
\begin{equation}\label{eq skew fraction}
\alpha={\left(\varphi_{C\mu\mu}^{(r)} \ \varphi_{C\sigma\sigma}^{(r)}\right)}^{\tau} \ ,
\end{equation}
which is just a simple power-law ansatz, guaranteeing that when $\varphi_{C\mu\mu}^{(r)} = 0$ or $\varphi_{C\sigma\sigma}^{(r)} = 0$ the global skewness comes from the local one alone, i.e. $\alpha = 0$, without introducing further parameters.
This is somehow required by the fact that when $\varphi_{C\mu\mu}^{(r)} = 0$ or $\varphi_{C\sigma\sigma}^{(r)} = 0$ the covariance is not well defined\footnote{In general, Eq. (\ref{eq skew fraction}) would require more investigation but, in practice, hereafter we model $\varphi_{C\mu\mu}^{(r)}$ and $\varphi_{C\sigma\sigma}^{(r)}$ as constant functions and the relation between $\alpha$ and $\tau$ becomes trivial.}.
The same argument does not apply for $\varphi_{C\mu\mu}^{(t)}$ and $\varphi_{C\sigma\sigma}^{(t)}$ because, for $\mu_\theta=0$, the skewness disappears by symmetry.
Clearly, in terms of pros and cons, this second approach is exactly the opposite of the first one. 
\end{enumerate}

Having tested both of the above solutions, we implement approach (ii).
The reason behind this choice is that the profile of the redshift-space correlation function obtained via approach (i) is affected by the presence of wiggles on small scales, which might induce an artificial scale dependence, e.g. when fitting for cosmological parameters.   
Likely, these undesired features are a direct consequence of the non-smooth behaviour discussed above.
As for the amplitude of the local skewness, we can roughly estimate $\gamma_L \in (-0.3,0)$.
Note, however, that this is an indirect measurement, obtained by assuming the model introduced in Sec. \ref{sec simple}, and as such it should be intended as a consistency test to ensure that the deviations from local Gaussianty are not too large.

\subsection{Simplest possible ansatz}\label{sec simple}

For a model to be useful it is important to keep it as simple as possible (but no simpler).
With this in mind, we discuss here the simplest possible ansatz for the parameters $\kappa_\mu^{(r)}$, $\kappa_\sigma^{(r)}$, $\kappa_\mu^{(t)}$, $\kappa_\sigma^{(t)}$, $r_g$ and $\tau$.
\begin{enumerate}
\item
Although the univariate-Gaussian assumption has been proved successful in describing the large scale behaviour of massive halos from N-body simulations, we know that the true velocity PDF never really reaches the Gaussian limit \citep[e.g.][]{scoccimarro2004b}.
In fact, even in linear theory, the multivariate Gaussian joint distribution of density and velocity field does not yield a Gaussian line-of-sight pairwise velocity PDF \citep{fisher1995}.
Furthermore, we expect the higher order moments of the velocity PDF to become important only on relatively small scales where the correlation function is steeper, see e.g. Eq. (15) in Paper I, or, in other words, we expect the shape of the velocity PDF not to be particularly relevant on large scales.
This suggest that we adopt $r_g=+\infty$.
\item
A relevant part of the global skewness is due to the covariance $C_{\mu\sigma}$ between local infall and velocity dispersion (Paper I and \citealt{tinker2007}).
We have just shown that, when GG is assumed, the maximum efficiency in converting the covariance into skewness is obtained for $\varphi_{C\mu\mu} = \varphi_{C\sigma\sigma}=1/3$.
In general, the skewness reaches its maximum for $\mu_\theta=1$ and disappears for $\mu_\theta=0$.
This suggests that we adopt $\kappa_\mu^{(r)}=\kappa_\sigma^{(r)}=1/3$.
\item
Similarly, the value of $\tau$ must be small enough to be compatible with the general picture in which the skewness is largely sourced by the covariance.
On the other hand, it cannot be zero because of the skewness problem described in Sec. \ref{sec skew problem}.
Based on our measurements, in the most extreme cases, corresponding to high-redshift DM and low-redshift small-mass halos, the skewness can exceed the upper limit given by GG of $\sim 40\%$.
From Eqs.~(\ref{eq skew frac})~and~(\ref{eq skew fraction}) it is easy to see that this missing skewness can be obtained by setting\footnote{In this final model, $\tau = 1/4$ corresponds to $\alpha \sim 0.6$.
As a reference, $\tau = 0$ corresponds to $\alpha = 1$ (i.e. GG is recovered), whereas $\tau > 1$ corresponds to $\alpha \sim 0$ (i.e. the global skewness is sourced by the local skewness alone).
We tested the performance of the model for $0.4 \lesssim \alpha \lesssim 0.8$, and we concluded that, within this range, variations in $\alpha$ can be effectively absorbed in small changes of the free parameters $\kappa_\mu^{(t)}$ and $\kappa_\sigma^{(r)}$.} $\tau=1/4$.
\end{enumerate}

With these ansatze, our model only depends on the first three moments of the velocity PDF, $m^{(1)}$, $c^{(2)}$ and $c^{(3)}$, plus two free parameters, $\kappa_\mu^{(t)}$, $\kappa_\sigma^{(t)}$.
We show in Sec. \ref{sec sim} that this model gives a good description of the redshift-space clustering.
It should nonetheless be said that, if we are interested in the true shape of the velocity PDF, e.g. when dealing with direct measurements of the velocity field \citep[e.g.][]{springob2007, tully2013}, the above assumptions should be relaxed.

\subsection{Model}

We provide a brief summary of our methodology for modelling the redfshift-space clustering:
\begin{enumerate}
\item
The first three velocity moments can be decomposed in radial and tangential components $m^{(1)}_r$, $c^{(2)}_r$, $c^{(2)}_t$, $c^{(3)}_r$ and $c^{(3)}_t$, which depend on the real-space separation $r$, but not on $\mu_\theta$, see App.~\ref{app moments}.
We assume that these quantities can be predicted theoretically as a function of cosmological parameters.
\item
We evaluate the scale-dependent GQG parameters as
\begin{align}
M_\mu &= m^{(1)}_r(r) \ \mu_\theta \\
C_{\mu\mu} &=  \frac{1}{3}c^{(2)}_r(r) \ \mu_\theta^2 + \kappa_\mu^{(t)} c^{(2)}_t(r) \ \left(1-\mu_\theta^2\right) \\
C_{\sigma\sigma} &=  \frac{1}{3}c^{(2)}_r(r) \ \mu_\theta^2 + \kappa_\sigma^{(t)} c^{(2)}_t(r) \ \left(1-\mu_\theta^2\right) \\
{M_\sigma}^2 &= c^{(2)}_r(r) \ \mu_\theta^2 + c^{(2)}_t(r) \ (1-\mu_\theta^2) - C_{\mu\mu} - C_{\sigma\sigma} \\
C_{\mu\sigma} &=  \frac{1}{\sqrt{3}}  \frac{\left[c^{(3)}_r(r) \ \mu_\theta^2 + c^{(3)}_t(r) \ \left(1-\mu_\theta^2\right) \right] \mu_\theta}{6M_\sigma} \\
\gamma_L &= \left(1-\frac{1}{\sqrt{3}}\right) \ \frac{\left[c^{(3)}_r(r) \ \mu_\theta^2 + c^{(3)}_t(r) \ \left(1-\mu_\theta^2\right) \right] \mu_\theta}{M_\sigma ({M_\sigma}^2 + 3C_{\sigma\sigma})} \ ,
\end{align}
where $\kappa_\mu^{(t)}$, $\kappa_\sigma^{(t)} \in (0,1)$, with $\kappa_\mu^{(t)} + \kappa_\sigma^{(t)} \le 1$, are scale-independent dimensionless parameters, which, in the simplest scenario, can be used as nuissance parameters or tuned to simulations.
\item
We use the GQG parameters to compute the scale-dependent velocity distribution, $\mP$, via Eq.~(\ref{eq GQG1dim}).
The procedure is self consistent, i.e. the second moment of the so obtained distribution is exactly $c^{(2)} = c^{(2)}_r(r) \ \mu_\theta^2 + c^{(2)}_t(r) \ \left(1-\mu_\theta^2\right)$, regardless of the amplitude of $\kappa_\mu^{(t)}$ and $\kappa_\sigma^{(t)}$, and similarly for $m^{(1)}$ and $c^{(3)}$.
\item
We use $\mP$ and the real-space correlation function $\xi_R$ to obtain the redshift-space correlation function $\xi_S$ via Eq.~(\ref{eq streaming}), where $\xi_R$ is assumed to be predicted by theory or measured from data \citep[e.g.][]{saunders1992}.
\end{enumerate}

The above equations refer to the ``simplest possible ansatz'' discussed in Sec. \ref{sec simple}, but the generalisation to a more complex scenario is straightforward.

\section{Comparison with simulations}\label{sec sim}

For our investigation we use the data from the MultiDark MDR1 run \citep{prada2012}, which follows the dynamics of $2048^3$ particles over a cubical volume of ${(1000 h^{-1} \text{Mpc})}^3$. 
The set of cosmological parameters assumed for this simulation is compatible with WMAP5 and WMAP7 data, $\{\Omega_m, \Omega_{\Lambda}, \Omega_b, \sigma_8, n_s\} = \{0.27, 0.73, 0.047, 0.82, 0.95\}$.
We consider three different redshifts, $z=0$, $z=0.5$ and $z=1$.
For each redshift we consider dark matter particles and two mass-selected halo catalogues, $10^{12} < (M/M_\odot) < 10^{13}$ and $M > 10^{13} M_\odot$.
The halos are identified via a friend-of-friend algorithm, with linking length 0.17.  

Since we assume that the fist three moments $m^{(1)}$, $c^{(2)}$ and $c^{(3)}$ are known, as well as the real-space correlation function $\xi_R$, we directly measure them from the simulation.
We also estimate from the simulation the overall line-of-sight pairwise velocity PDF $\mP$, which we use as a reference for model comparison. 
The procedures adopted for all these measurements are reported in App.~\ref{app moments}. 

\begin{figure*}
 \begin{center}
   \includegraphics[width=18cm]{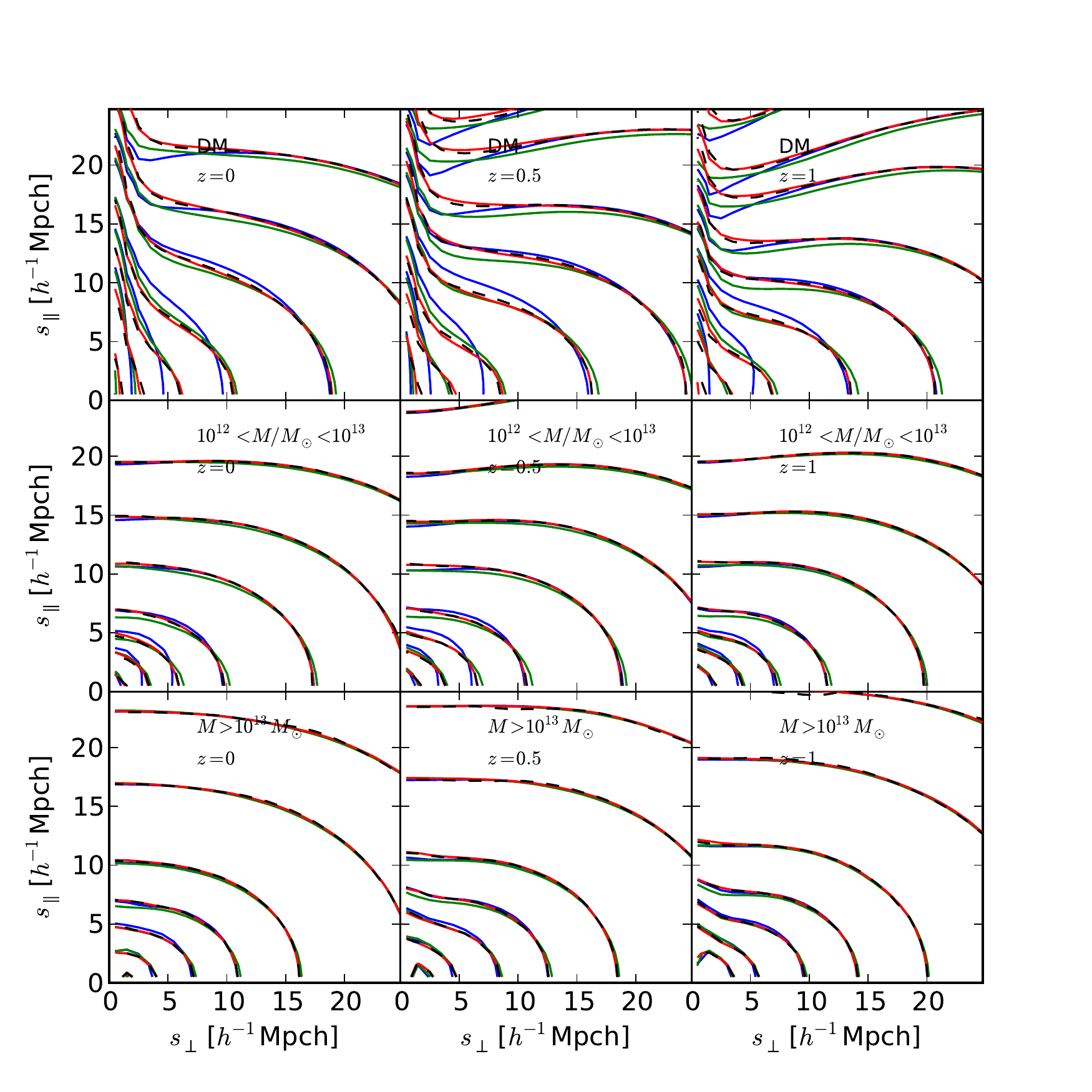}
    \caption{Redshift-space correlation function as a function of the redshift-space separation parallel and perpendicular to the line of sight, $s_\parallel$ and $s_\perp$, respectively, for different tracers and redshifts as labeled in the figure.
The iso-correlation contours are obtained via the streaming model with different assumptions for the line-of-sight velocity PDF, specifically:
direct measure from the simulations, black dashed;
direct measure of the first three moments from the simulations plus GQG assumption, red solid;
direct measure of the first three moments from the simulations plus Edgeworth expansion, blue solid;
direct measure of the first two moments from the simulations plus univariate Gaussian assumption, green solid.}
  \label{fig xis_multi}
 \end{center}
\end{figure*}

In Fig. \ref{fig xis_multi} we present the redshift-space 2-dimensional correlation function $\xi_S(s_\perp, s_\parallel)$ obtained via the streaming model, Eq.~(\ref{eq streaming}), with various assumptions, and compare these with the measured velocity PDF (see App. \ref{app sys error} for the correspondent fractional deviations).
The lines represent:
\begin{enumerate}
\item
direct measure of the velocity PDF from the simulations, black dashed;
\item
direct measure of the first three moments from the simulations plus GQG assumption for the velocity PDF, red solid;
\item
direct measure of the first three moments from the simulations plus Edgeworth expansion for the velocity PDF, blue solid;
\item
direct measure of the first two moments from the simulations plus univariate Gaussian assumption for the velocity PDF, green solid.
\end{enumerate}
Since for each of the above we use the same "true" real-space correlation function $\xi_R(r)$, any difference in the corresponding $\xi_S(s_\perp, s_\parallel)$ can be attributed to the impact of different assumptions on the shape of the velocity PDF.
As discussed in Sec. \ref{sec strategy}, the GQG distribution requires additional knowledge of the functions $\varphi$.
These latter, under the simplest possible ansatz, can be parametrised by $\kappa_\mu^{(t)}$ and $\kappa_\sigma^{(t)}$, Sec.~\ref{sec simple}.
We fit these parameters to simulations.
Since on large scales all the models perform well, here we focus on small-to-intermediate scales.
As can be seen from the figure, for any tracer and redshift considered, the GQG prescription improves on the Edgeworth streaming model (ESM), which in turn improves on the Gaussian streaming model.
This is somewhat expected, given the different number of degrees of freedom of the different models (nonetheless, we note that on the smallest scales the GSM seems to perform slightly better than the ESM even though it has fewer degrees of freedom).
Specifically, the smaller the mass of the tracer the larger the improvement provided by GQG with respect to ESM and GSM, which is also expected, since the velocity PDF becomes less and less Gaussian going from massive to less-massive halos and then to DM.

\begin{figure*}
 \begin{center}
  \includegraphics[width=18cm]{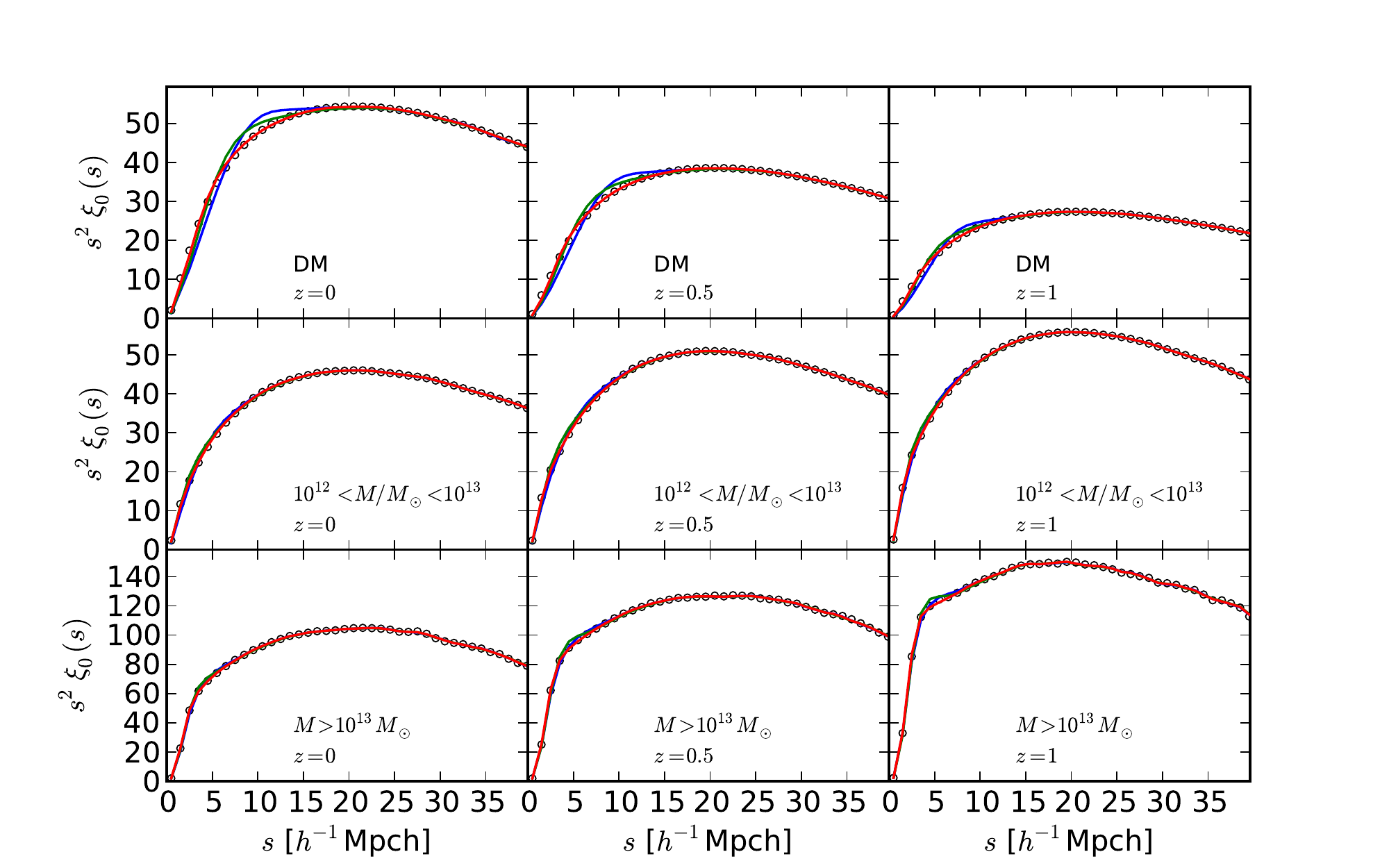}
  \caption{Legendre monopole of the redshift-space correlation function $\xi_0(s)$ as a function of the redshift-space separation $s$, for different tracers and redshifts as labeled in the figure.
  Following common practice, on the $y$ axis we report $s^2 \xi_0(s)$, in order to help in the visualisation of the large-scale behaviour.
  The lines correspond to the same models as in Fig. \ref{fig xis_multi}, with the same colour coding, except for the direct measurement of the velocity PDF, which is here represented by open circles.}
  \label{fig mpoles0}
 \end{center}
\end{figure*}

\begin{figure*}
 \begin{center}
  \includegraphics[width=18cm]{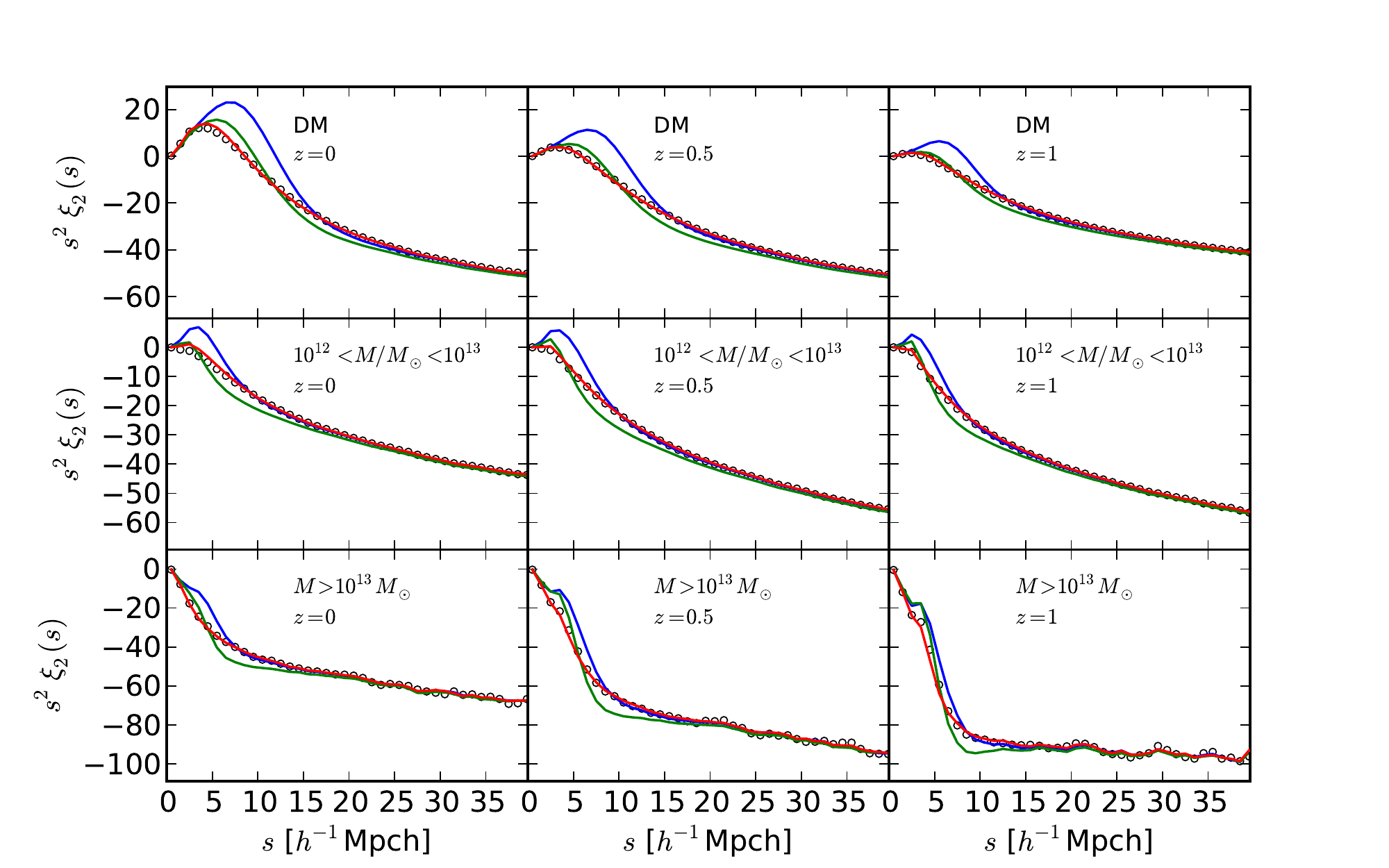}
  \caption{Same as Fig. \ref{fig mpoles0} but for the quadrupole of the redshift-space correlation function $\xi_2(s)$.}
  \label{fig mpoles2}
 \end{center}
\end{figure*}

\begin{figure*}
 \begin{center}
  \includegraphics[width=18cm]{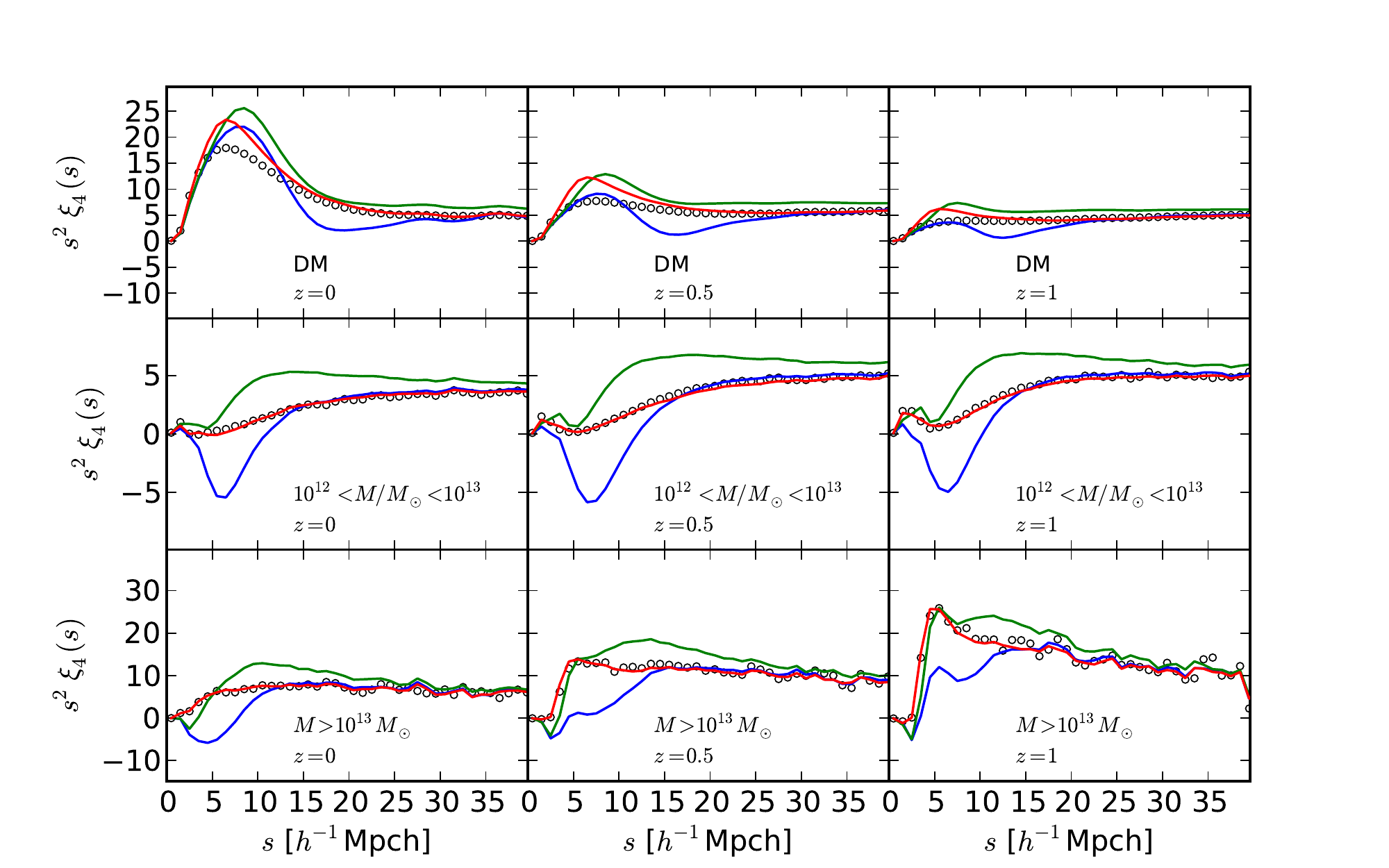}
  \caption{Same as Fig. \ref{fig mpoles0} but for the hexadecapole of the redshift-space correlation function $\xi_4(s)$.}
  \label{fig mpoles4}
 \end{center}
\end{figure*}

In Figs.~\ref{fig mpoles0}, \ref{fig mpoles2} and \ref{fig mpoles4} we plot the first three even Legendre multipoles of the redshift-space correlation function, namely the monopole $\xi_0(s)$,  the quadrupole $\xi_2(s)$ and the hexadecapole $\xi_4(s)$.
In general, Legendre multipoles are preferred with respect to the full 2-dimensional correlation function when fitting models to the data because it is easier to estimate the correspondent covariance matrix.   
Monopole and quadrupole moments have been recently used for estimation of the cosmological parameters via the GSM \citep[e.g.][]{samushia2014}.
The monopole, Fig.~\ref{fig mpoles0}, is quite accurate for all the three models considered, with a small deviation of ESM and GSM from the expected amplitude on small scales in the DM case.
This small-scale inaccuracy becomes more important when we consider the quadrupole, Fig.~\ref{fig mpoles2}.
Specifically, the ESM  is biased for scales $\lesssim 10-15 h^{-1}$Mpc, depending on tracer and redshift, whilst the GSM starts failing on $\sim 10 h^{-1}$Mpc larger scales\footnote{With respect to a similar consistency test of the GSM reported in figure 6 of \citet{reid2011}, we note some discrepancy in the small-scale behaviour, especially for the quadrupole.
The origin of this discrepancy is not clear, however the overall message of Reid and White's work, i.e. the GSM is few-percent precise on scales $\gtrsim 30 h^{-1}$Mpc for the monopole and quadruple of standard halo populations, is compatible with our results.}.
On the other hand, as already noted, on the smallest scales the deviation from the expected amplitude is more severe for the ESM.
The GQG distribution is instead in good agreement with the direct measurements on all scales.
A similar behaviour is found for the hexadecapole, Fig.~\ref{fig mpoles4}.
In this case the ESM fails on scales $\lesssim 15-30 h^{-1}$Mpc, depending on tracer and redshift, whilst the GSM is biased on all the scales considered.
The GQG prescription recovers the correct amplitude on all scales, except for a deviation on small scales in the DM case.
We attribute this deviation to the simplistic form we have assumed for the functions $\varphi$.
Very likely, it would be possible to improve on this by allowing for more general functional forms (see appendix \ref{sec smallscale} for a more realistic description of the small-scale behaviour), nonetheless, since the issue appears in the DM case only, in this work we prefer not to further complicate the model. 

\begin{figure}[h!]
 \begin{center}
  \includegraphics[width=9cm]{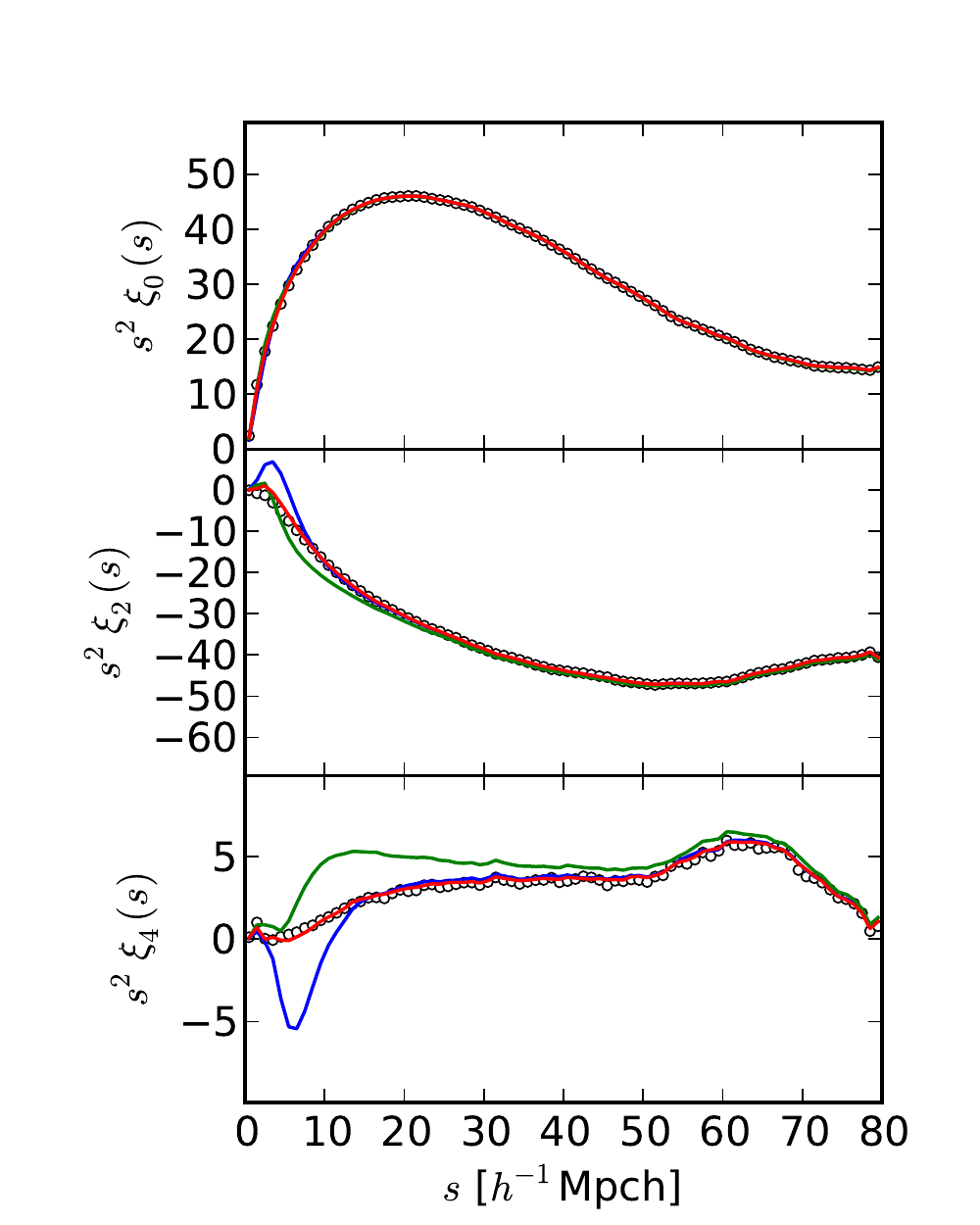}
  \caption{Legendre monopole, quadrupole and hexadecapole of the redshift-space correlation function, for the halo catalogue $10^{12} < (M/M_\odot) < 10^{13}$ at $z=0$, on a large separation range, $0< s < 80h^{-1}$Mpc.
  The lines correspond to the same models as in Fig. \ref{fig mpoles0}, with the same colour coding.}
  \label{fig mpoles largescale}
 \end{center}
\end{figure}

For completeness, in Fig. \ref{fig mpoles largescale} we show the multipoles of the correlation function for the $10^{12} < (M/M_\odot) < 10^{13}$ halo catalogue over a broader range of separations.
As anticipated, on large scales all the three models tend to match the expected amplitude.  
We note however that the GSM is wrong even on moderate scales for the hexadecapole.

\section{Discussion and conclusions}\label{sec discu}

It is well known that a percent-level understanding of the anisotropy
of the redshift-space galaxy clustering is needed to accurately recover cosmological information from the RSD signal in order to shed light on the issue of dark energy vs. modified gravity. 
From a statistical point of view, the source of the anisotropy is the galaxy line-of-sight pairwise velocity distribution.
It is therefore important to adopt a realistic functional form for this velocity PDF when fitting models to the data.   
To this purpose, in paper I we introduced the GG prescription for the velocity PDF.
In this work we have continued the development of this model by making explicit the dependence of the GG distribution on quantities predictable by theory, namely its first three moments, and extending it to the more general concept of GQG.
To keep the model as simple as possible, we have proposed an ansatz with two free dimensionless parameters that describe how infall velocity and velocity dispersion vary when moving from one place to another in our Universe.
Since their interpretation is clear, these parameters can be theoretically predicted or, assuming a more pragmatic approach, tuned to simulations or used as nuisance parameters.
State-of-the-art PT has proven successful in predicting the large-scale behaviour of the velocity PDF and the correspondent monopole and quadrupole of the redshift-space correlation function \citep[e.g.][]{reid2011, wang2014}, at least for massive halos, $M \sim 10^{13}M_\odot$.  
Unfortunately, by definition, any PT breaks down for small separations.
Consequently, alternative approaches have been suggested in the literature, spanning from purely theoretical \citep[e.g.][]{sheth1996} to hybrid techniques in which N-body simulations plus a halo occupation distribution (HOD) are employed to deal with the issue of non linearities \citep[e.g.][]{tinker2007, reid2014}.
One of the main results from our work is to provide a framework in which perturbation and small-scale theories are smoothly joined, so that all available RSD information can be coherently extracted from redshift surveys. 
A fundamental requirement for a redshift-space model is that it must be precise on all scales interest, and it should inform the user of the scales on which the model can be trusted.
We have compared to N-body simulations the well know GSM \citep{reid2011}, the more recent ESM \citep{uhlemann2015} and the GQG prescription over a broad range of separations, from $0$ to $80h^{-1}$Mpc.
Different redshifts, from $z=0$ to $z=1$, and different tracers, namely DM particles and two mass-selected catalogues of DM halos, have been considered.   
We have concluded that, among the three, QGQ is the only model capable of providing a precise redshift-space correlation function on scales down to $\sim 5h^{-1}$Mpc over the range of redshifts covered by future surveys.
Keeping in mind that the range of validity of the models depends on tracer, redshift and order of the Legendre multipoles we are interested in, for finiteness, we can say that all the models converge to the aspected amplitude on scales $\gtrsim 30h^{-1}$Mpc, at least for multipole and quadrupole.
Since these scales roughly coincide with the range of validity of state-of-the-art PTs, if we rely only on PT and if we are not interested in higher order multipoles, the most natural choice is the simplest model among the three, i.e. the GSM.   
As for the ESM, we have found it to be unbiased down to smaller scales and for higher order multipoles than the GSM, thus confirming the results by \citet{uhlemann2015}, but, on the other hand, it seems to behave even worse than the GSM on the smallest scales. 
We can therefore think of it as a natural extension of the GSM in the perspective of further PT developments.  
In particular, a better prediction of the third moment of the velocity
PDF is required before the ESM can be applied to data on smaller scales.
Formally, the same argument holds for the GQG model, nonetheless, since this latter is meant to include nonlinear scales, it could be possible to obtain a prediction for the third moment by interpolating between (very) small and (very) large scales.
More precisely, as shown in the lower right panel of Fig. \ref{fig moms multi}, the functions $c^{(3)}_t$ and $c^{(3)}_r$, which fully characterise the third moment, are peaked at $r \lesssim 10h^{-1}$Mpc.
By adopting a model for the small-scale limit that includes those separation, most likely using simulations in a similar way to that proposed in \citet{reid2014}, we would then be able to interpolate between these peaks and their large-scale limit, which is trivially 0.

For the above reasons, we have not tested here the performance of the of the models in recovering cosmological parameters, the growth rate $f$ in particular.
This important topic will be explored in a further work in which a prescription for the small-scale limit will be discussed.   
Also left for further work is an extensive test of the model on realistic mock galaxy catalogues, which very likely will give results somewhere in between those obtained for DM and halos.  

Another interesting question to be answered is whether the GQG distribution can play a role in the interpretation of the data coming from direct measurements of the velocity of galaxies \citep[e.g.][]{springob2007, tully2013}, or, conversely, whether these data can be helpful in tuning the GQG parameters.  

The moments of the velocity PDF on small scales are extremely sensitive to deviations from GR \citep[e.g.][]{fontanot2013, hellwing2014}.
Constraining these quantities is therefore of particular interest in understanding gravity. 
Although we have tested our model against $\Lambda$CDM simulations
only, at no stage of its derivation have we assumed GR.
Further investigation is clearly needed into this topic, but we do not see any obvious reason for the model not to be compatible with modified-gravity velocity PDFs and clustering. 

Similarly, we do not expect baryonic physics to invalidate the GQG description, but, obviously, taking into account the impact of baryons makes the theoretical prediction of the very small scales more challenging.    

Finally, we note that we have defined and analysed a very general probability distribution function, the GG distribution, which could prove useful in completely different fields.
As a generalisation, we have also introduced the GQG distributions, which is formally a pseudo distribution, since for extreme values of the local skewness it can assume negative amplitude.
It is nonetheless important to note that, at variance with what we have found for the standard Edgeworth expansion, in our measurements this unphysical situation never occurs.

\section*{Acknowledgements}

DB and WJP are grateful for support from the European Research Council
through the grant 614030 ``Darksurvey''.  WJP is also grateful for
support from the UK Science and Technology Facilities Research Council
through the grant ST/I001204/1.
JB acknowledges support of the European Research Council through the Darklight ERC Advanced Research Grant (\#291521). 




\bibliographystyle{mnras}
\bibliography{./biblio_db}




\appendix

\section{Measurements of velocity PDF, moments and correlation function from simulations}\label{app moments}

Ignoring wide-angle effects, the line-of-sight pairwise velocity distribution $\mP(v_\parallel |r_\perp,r_\parallel)$ is obtained by projecting along the line of sight the 2-dimensional pairwise velocity distribution $\mP_r(v_r,v_t|r)$.
This latter is the joint distribution of the parallel ($v_r$) and perpendicular ($v_t$) components of the pairwise velocity with respect to the pair separation ${\bf r}$.
Due to isotropy it depends only on the length $r$ of the separation vector.
Although measurements of $\mP$ and $\mP_r$ are formally equivalent, we prefer to adopt the second approach since it allows us to take advantage of all possible symmetries, thus minimising statistical noise and cosmic variance (in essence, we do not need to choose a line of sight).

Similarly, the moments of the pairwise-velocity PDF can be decomposed as follows \citep[e.g.][]{uhlemann2015},
\begin{align}
m^{(1)}(r,\mu_\theta) &= m^{(1)}_r(r) \ \mu_\theta \label{eq dec m1} \\
c^{(2)}(r,\mu_\theta) &= c^{(2)}_r(r) \ {\mu_\theta}^2 + c^{(2)}_t(r) \left(1-{\mu_\theta}^2 \right) \label{eq dec c2}\\
c^{(3)}(r,\mu_\theta) &= \left[c^{(3)}_r(r) \ {\mu_\theta}^2 + c^{(3)}_t(r) \left(1-{\mu_\theta}^2 \right)\right]\mu_\theta \label{eq dec c3} \ ,
\end{align}
where $r=\sqrt{{r_\perp}^2+{r_\parallel}^2}$, $\mu_\theta=r_\parallel/r$ and we have used the fact that,
because of isotropy, the only non-vanishing correlators between the radial and tangential component of the pairwise velocity are those involving even powers (i.e. the modulus) of the tangential component.
Here, for self consistency and to minimise the statistical noise, it is convenient to follow a scheme that is somehow opposite to what we do for the PDF: from $\mP$ we measure the left-hand term of these equations, but our model requires as an input the radial-dependent functions on the right-hand side, which is what is usually predicted in PT. 
We then need to invert this set of equations. 
From Eq. (\ref{eq dec m1}) we get
\begin{equation}\label{eq m1pa}
m^{(1)}_r(r) = \frac{1}{\Delta\mu_\theta}  \int_{\Delta\mu_\theta} d \mu_\theta \ \frac{m^{(1)}(r,\mu_\theta)}{\mu_\theta} \ ,
\end{equation}
where the integral can in principle be performed over any arbitrary interval $\Delta\mu_\theta=\mu_\theta^{max}-\mu_\theta^{min}$, with $0<\mu_\theta^{min}<\mu_\theta^{max}<1$.
Similarly, Eq. (\ref{eq dec c2}) yelds
\begin{align}
c^{(2)}_r(r) =& \ \frac{1}{\Delta\mu_\theta} \int_{\Delta\mu_\theta} d \mu_\theta \left[ \frac{2}{3 \mu_\theta^2  - 1} \ c^{(2)}(r,\mu_\theta) \right. \nonumber \\
&- \left. \frac{2(1-{\mu_\theta}^2)}{3 {\mu_\theta}^2  - 1} \ \bar{c}^{(2)}(r)\right] \\
c^{(2)}_t(r) =& \ \frac{1}{\Delta\mu_\theta} \int_{\Delta\mu_\theta} d \mu_\theta \left[ \frac{1}{1 - 3\mu_\theta^2} \ c^{(2)}(r,\mu_\theta) \right. \nonumber \\
&- \left. \frac{3\mu_\theta^2}{1 - 3 \mu_\theta^2} \ \bar{c}^{(2)}(r)\right] \ ,
\end{align}
where we have defined
\begin{equation}
\bar{c}^{(2)}(r) = \int_0^1 d\mu_\theta \ c^{(2)}(r,\mu_\theta) \ .
\end{equation}
Finally,  from Eq. (\ref{eq dec c3}) we obtain
\begin{align}
c^{(3)}_r(r) =& \ \frac{1}{\Delta\mu_\theta} \int_{\Delta\mu_\theta} d \mu_\theta \left[ \frac{1}{\mu_\theta(2\mu_\theta^2  - 1)} \ c^{(3)}(r,\mu_\theta) \right.\nonumber \\
 &- \left. \frac{4(1-\mu_\theta^2)}{2 \mu_\theta^2  - 1} \ \bar{c}^{(3)}(r)\right] \\
c^{(3)}_t(r) =& \ \frac{1}{\Delta\mu_\theta} \int_{\Delta\mu_\theta} d \mu_\theta \left[ \frac{1}{\mu_\theta(1 - 2\mu_\theta^2)} \ c^{(3)}(r,\mu_\theta) \right. \nonumber \\
&- \left. \frac{4\mu_\theta^2}{1 - 2 \mu_\theta^2} \ \bar{c}^{(3)}(r)\right] \ ,
\end{align}
where
\begin{equation}
\bar{c}^{(3)}(r) = \int_0^1 d\mu_\theta \ c^{(3)}(r,\mu_\theta) \ .
\end{equation}
Clearly, the larger $\Delta \mu_\theta$ the more information we include in our analysis, nonetheless two potential issues have to be considered.
\begin{enumerate}
\item
For $\mu_\theta=1$ the integrals might diverge.
This problem is naturally solved by the fact that the moments are measured in bins of $\mu_\theta$, which means that the largest available $\mu_\theta^{max}$ is always smaller than~1.   
\item
Since the odd moments vanish for $\mu_\theta \rightarrow 0$ [Eqs. (\ref{eq dec m1}) and (\ref{eq dec c3})], including small values of $\mu_\theta$ in our analysis only add instability. For these moments we then safely adopt $\mu_\theta^{min} = 0.5$.
\end{enumerate}

\begin{figure*}
 \begin{center}
   \includegraphics[width=18cm]{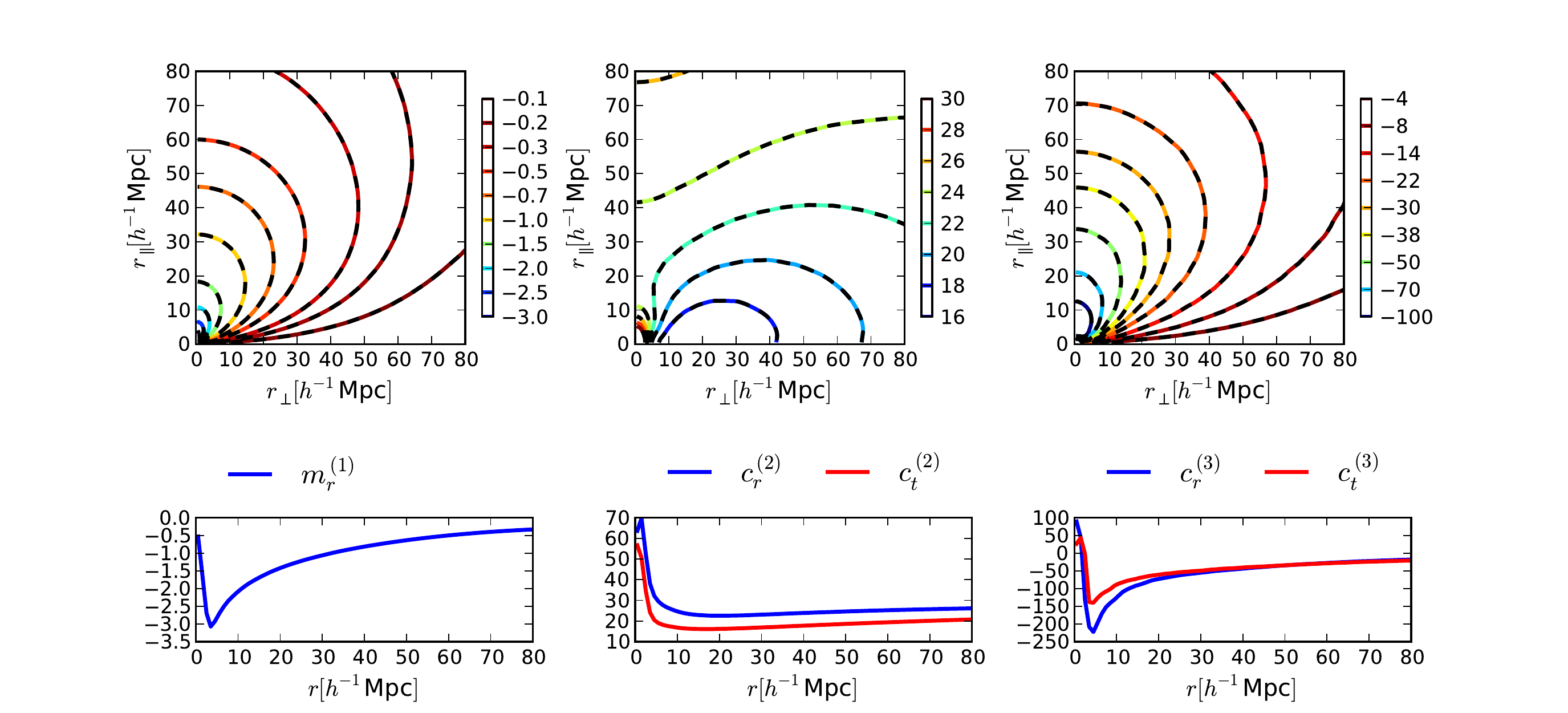}
    \caption{Moments of the line-of-sight pairwise velocity distribution for the DM catalogue at $z=0$ (the velocity is measured in units of length via the standard $\mathcal{H}^{-1}$ rescaling).
    Upper panels, from left to right: first moment $m^{(1)}$, second
    central moment $c^{(2)}$ and third central moment $c^{(3)}$,
    presented using iso-amplitude contours as a function of the real-space separation parallel and perpendicular to the line of sight, $r_\parallel$ and $r_\perp$, respectively.
Different lines represents:  direct measurements of the moments from simulations, solid coloured, reconstruction of the moments via Eqs. (\ref{eq dec m1}), (\ref{eq dec c2}) and (\ref{eq dec c3}).
Lower panels: measurements of the parallel and perpendicular (with respect to the real-space separation vector ${\bf r}$) component of the moments as a function of $r$, as labelled in the figure.
These functions are used as input for computing the dashed lines in the correspondent upper panel.}
  \label{fig moms multi}
 \end{center}
\end{figure*}

In the left upper panel of Fig.~\ref{fig moms multi} we compare the direct measurement from $\mP$ of the first moment $m^{(1)}$ (solid lines) with that obtained by estimating $m_r^{(1)}$  via Eq. (\ref{eq m1pa}) and than multiplying by $\mu_\theta$ (dashed lines).
In other words we test the validity of our approach by comparing left- with right-hand side of Eq. (\ref{eq dec m1}).
We do the same for the less trivial measurements of $c^{(2)}$ and $c^{(3)}$, central and right upper panel, respectively.
Only contours from the DM catalogue at $z=0$ are shown but all tracers and redshift considered yield similar results.
We also report in the lower panels the underling decomposition of the moments.    
Given the good match seen in the figures, we can conclude that our procedure to decompose the moments works properly and will not introduce any kind of bias in our final results.

For the estimation of correlation function we adopt the natural estimator $\xi = \frac{DD}{RR}-1$, where $DD$ and $RR$ represent the number of data and random pairs at a given separation, respectively.
This is the most natural choice when dealing with periodical boxes, in which there are no border effects and $RR$ can be computed analytically.

For all the measurements in this work we adopt linear bins of $1h^{-1}$Mpc size [note that, since we use the standard $\mathcal{H}^{-1}$ rescaling \citep[see e.g.][]{scoccimarro2004b}, the velocities are measured in unit of length].

\section{Details on the moment generating function}\label{app momgenfun}

\begin{figure}
 \begin{center}
   \includegraphics[width=9cm]{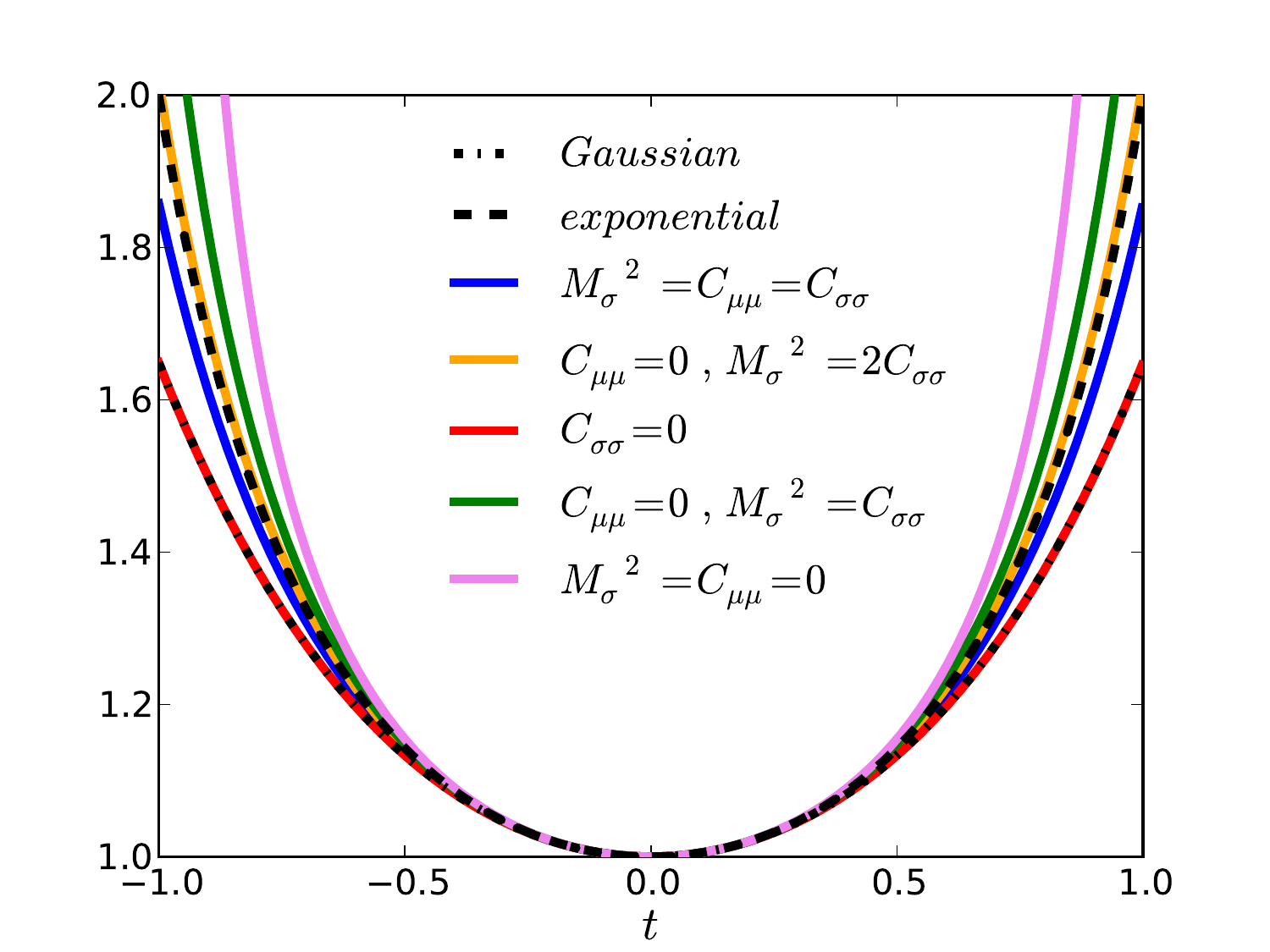}
    \caption{Moment generating functions of the GG distribution for different combinations of the parameters $M_\sigma$, $C_{\mu\mu}$ and $C_{\sigma\sigma}$ as labelled in the figure, solid coloured.
    Only the zero-skewness case is considered, i.e. $C_{\mu\sigma}=0$.
    For comparison, we also show the moment generating function of Gaussian and exponential distributions, black dot-dashed and black dashed, respectively.
     All the curves are standardised, i.e zero mean and unit variance.}
  \label{fig GGgeneratingfuns}
 \end{center}
\end{figure}

For the sake of completeness, in table \ref{tab momgenfun} we report the MGF of the GG distribution for a few specific combinations of the parameter set $\{M_\mu, M_\sigma, C_{\mu\mu}, C_{\sigma\sigma}, C_{\mu\sigma}\}$.
Specifically, from top to bottom we show:
\begin{enumerate}
\item
The MGF of the full distribution.
\item
The zero-skewness limit, $C_{\mu\sigma}=0$. This is the limit of the GG distribution for $\mu_\theta \rightarrow 0$ where the skewness disappears by symmetry.   
\item
The maximum-skewness limit, ${M_{\sigma}^2}=C_{\mu\mu}=C_{\sigma\sigma}$. Since, as shown in Sec. \ref{sec skew problem}, for this combination of the parameters the conversion of covariance in skewness is maximised, we assumed this limit for $\mu_\theta \rightarrow 1$. Note however that, in order to match simulations, we have to correct for the skewness by using GQG, Sec.~\ref{sec GQG}.
\item
The Gaussian limit, $C_{\sigma\sigma=0}$. This limit has been discussed in Sec. \ref{sec GG}. When the further condition $C_{\mu\mu=0}$ is added, we obtain a very natural large scale limit.
\item
The small-scale limit, $C_{\mu\mu}=C_{\mu\sigma}=0$. As discussed in Sec. \ref{sec GG} and, more extensively, in App. \ref{sec smallscale}, this is the behaviour we expect at very small separations, where the infall velocity disappears.
\item
The quasi-exponential limit, $C_{\mu\mu}=C_{\mu\sigma}=0$ and ${M_{\sigma}}^2=2C_{\sigma\sigma}$. The MGF of an exponential is $\exp(\mu t)/(1-\frac{1}{2}\sigma^2t^2)$, where $\mu$ is the mean and $\sigma^2$ the variance, which clearly differs from what is reported in the table. Nonetheless we show in Fig. \ref{fig GGgeneratingfuns} that for this combination of the parameters the MGFs of the two distributions behave in a very similar way.
\item
The combination $M_{\sigma}=C_{\mu\mu}=C_{\mu\sigma}=0$, which, formally, is another sub case of the small-scale limit.
Although we have not explicitly used such combinations it in this work, it is by itself interesting to see how simple becomes the MGF under this condition.
As far as we know, this do not correspond to the MGF of any common distribution but it helps us in showing how wide is the parameter space spanned by the GG distribution, see Fig.  \ref{fig GGgeneratingfuns}.   
\end{enumerate} 
\begin{table*}
\centering
\begin{tabular}{lll}
assumptions & moment generating function & n.f.p. \\[2.5mm]
\hline
\hline \\[-1.mm]
 - & $\frac{1}{\sqrt{1-t^2 C_{\sigma\sigma}}} \exp \left\{t M_\mu + \frac{1}{2} t^2 {M_\sigma}^2 + \frac{1/2}{1-t^2 C_{\sigma\sigma}} \left[ t^2C_{\mu\mu} + 2 t^3 C_{\mu\sigma} M_\sigma + t^4 \left({M_\sigma}^2 C_{\sigma\sigma} - \det{C}\right)\right] \right\}$ & 5 \\[5.mm]
$C_{\mu\sigma} = 0$ &  $\frac{1}{\sqrt{1-t^2 C_{\sigma\sigma}}} \exp \left\{t M_\mu + \frac{1}{2} t^2 {M_\sigma}^2 + \frac{1/2}{1-t^2 C_{\sigma\sigma}} \left[t^2 C_{\mu\mu} + t^4 C_{\sigma\sigma} \left({M_\sigma}^2 - C_{\mu\mu}\right)\right]  \right\}$ & 4 \\[5.mm]
${M_\sigma}^2 = C_{\mu\mu} =  C_{\sigma\sigma}$  & $\frac{1}{\sqrt{1-t^2 {M_\sigma}^2}} \exp \left\{t M_\mu  + \frac{1}{1-t^2 {M_\sigma}^2}\left[t^2 {M_\sigma}^2 + t^3 C_{\mu\sigma} M_\sigma + t^4\left(C_{\mu\sigma} - \frac{1}{2}{M_\sigma}^4\right)\right]\right\}$ & 3 \\[5.mm]
$C_{\sigma\sigma}  = C_{\mu\sigma} = 0$ &  $\exp \left[t M_\mu + \frac{1}{2} t^2 \left({M_\sigma}^2 + C_{\mu\mu} \right)\right]$ & 3 \\[5.mm]
$C_{\mu\mu}  = C_{\mu\sigma} = 0$ & $\frac{1}{\sqrt{1-t^2 C_{\sigma\sigma}}} \exp \left(t M_\mu + \frac{ \frac{1}{2} t^2 {M_\sigma}^2 }{1-t^2 C_{\sigma\sigma}} \right)$ & 3 \\[5.mm]
$C_{\mu\mu}  = C_{\mu\sigma} = 0 \quad, \quad {M_\sigma}^2 = 2C_{\sigma\sigma}$ & $\frac{1}{\sqrt{1-t^2 C_{\sigma\sigma}}} \exp \left(t M_\mu + \frac{t^2 C_{\sigma\sigma}}{1-t^2 C_{\sigma\sigma}} \right)$ & 2 \\[5.mm]
$M_\sigma = C_{\mu\mu} = C_{\mu\sigma} =0$ & $\frac{1}{\sqrt{1-t^2 C_{\sigma\sigma}}} \exp \left(tM_\mu\right)$  & 2
 \end{tabular}
\caption{Moment generating function of the GG distribution (central column) for different assumptions on the parameters $M_\mu$, $M_\sigma$, $C_{\mu\mu}$, $C_{\sigma\sigma}$ and $C_{\mu\sigma}$ (left column).
In the right column is reported the number of free parameters.} 
\label{tab momgenfun}
\end{table*}
In Fig. \ref{fig GGgeneratingfuns} we show the MGF of the GG distribution for the combination of parameters discussed above (coloured solid).
For comparison we also report the MGF of Gaussian and exponential distribution (black dot-dashed and black dashed, respectively).
All the functions have zero mean, unitary variance and zero skewness.
It is clear from the figure that the GG distribution efficiently covers the space between Gaussian and exponential distribution and beyond. 

\section{Bivariate distribution of $\mu$ and $\sqrt{\sigma}$ as an alternative way to allow for more skewness}\label{app alternative}

Most of the calculations presented in this paper can be easily extended to the case in which the jointly distributed variables are $\mu$ and $\sigma^{\frac{1}{2n}}$, with $n \in N$, rather than $\mu$ and $\sigma$.
Here we discuss the specific scenario in which $n=1$, i.e. $\mu$ and $\psi \equiv \sqrt{\sigma}$ are jointly distributed according to a bivariate Gaussian.
Since, as shown in the following, the resulting upper limit for the skewness of the velocity PDF is higher than that of a standard GG distribution, this approach potentially represents a viable alternative to GQG in solving the skewness issue, Sec. \ref{sec skew problem}. 

The integration over $\mu$ still gives Eq. (\ref{eq GG1dim}) but, obviously, with different expressions for $\mA$ and $\mK_i$,
\begin{align}
\mA^2 &= C_{\psi\psi} \psi^4 + C_{\mu\mu} C_{\psi\psi} - {C_{\mu\psi}}^2 \\
\mK_2 &=  C_{\psi\psi} \\
\mK_1 &= - 2 C_{\mu\psi} \left(\psi - M_\psi \right) \\
\mK_0 &=  \left(\psi^4 + C_{\mu\mu}\right) {\left(\psi - M_\psi \right)}^2 \ .
\end{align}
As for the first three moments,  we obtain
\begin{align}
m^{(1)} &= M_\mu \\ 
c^{(2)} &= {M_\psi}^4 + C_{\mu\mu} + 6 {M_\psi}^2 C_{\psi\psi} + 3 {C_{\psi\psi}}^2 \label{eq c2psi} \\
c^{(3)} &= 12 \left({M_\psi}^3 + 3 M_\psi C_{\psi\psi} \right) C_{\mu\psi} \label{eq c3psi} \ .
\end{align}
We can express Eq. (\ref{eq c3psi}) in terms of the correlation coefficient $\rho = C_{\mu\psi}/\sqrt{C_{\mu\mu} C_{\psi\psi}}$,
\begin{equation}\label{eq c3psirho}
c^{(3)} = 12 \rho \sqrt{{\left({M_\psi}^3 + 3 M_\psi C_{\psi\psi}\right)}^2 C_{\mu\mu} C_{\psi\psi}}
\end{equation}
In analogy to what we have done in Sec \ref{sec model}, we define
\begin{equation}
\varphi_{M\psi} \equiv \frac{{M_\psi}^2}{\sqrt{c^{(2)}}} \quad \varphi_{C\psi\psi} \equiv \sqrt{\frac{3}{c^{(2)}}} C_{\psi\psi} \quad \varphi_{C\mu\mu} \equiv \frac{C_{\mu\mu}}{c^{(2)}} \ ,
\end{equation}
for which holds the relation
\begin{equation}\label{eq consist rel psi}
\varphi_{C\mu\mu}  + \varphi_{M\psi} ^2 + 2 \sqrt{3} \varphi_{M\psi} \varphi_{C\psi\psi} + \varphi_{C\psi\psi}^2 = 1.
\end{equation}
We then rewrite Eq. (\ref{eq c3psirho}) as
\begin{equation}\label{eq skew lim psi}
c^{(3)} = 4 {\left(\sqrt{3} {c^{(2)}}\right)}^{3/2} \rho \left(\varphi_{M\psi} + \sqrt{3} \varphi_{C\psi\psi}\right) \sqrt{\varphi_{M\psi} \varphi_{C\psi\psi} \varphi_{C\mu\mu}} \ .
\end{equation}
Since by construction $|\rho|<1$, form Eqs. (\ref{eq consist rel psi}) and (\ref{eq skew lim psi}) we can assess the upper limit for the skewness $\gamma = \frac{c^{(3)}}{{c^{(2)}}^{\frac{3}{2}}}$.
Specifically, we obtain $|\gamma| \lesssim 1.85$, which is $\sim 60\%$ larger than what we get for a standard GG distribution, Sec. \ref{sec skew problem}.

\section{Small scale limit}\label{sec smallscale}

For very small separations, $r \rightarrow 0$, the velocity statistics is dominated by pairs inside virialized region, we therefore expect the local infall velocity to disappear, which implies $M_\mu = C_{\mu\mu} = 0$.
Note that the latter equality requires $C_{\mu\sigma} = 0$ as well.
By substituting in Eq. (\ref{eq GG1dim}) we find
\begin{align}
\label{eq smallscalegg}
\mP &= \int d \sigma \ \frac{\mW}{\sqrt{2 \pi \sigma^2}} \exp \left(-\frac{\vpa^2}{2 \sigma^2} \right)\ ,
\end{align}
where
\begin{align}
\mW \equiv \frac{1}{\sqrt{2 \pi C_{\sigma\sigma}}} \exp \left[-\frac{{(\sigma - {M_\sigma})}^2}{2 C_{\sigma\sigma}} \right]  \ .
\end{align}
Following the same reasoning behind Eqs. (\ref{eq gau1biv2gen}) and (\ref{eq bpm}), we define
\begin{align}\label{eq wpm}
\mW^\pm(\sigma) &\equiv \mW(-\sigma) + \mW(\sigma) \nonumber \\
&= \sqrt{\frac{2}{\pi C_{\sigma\sigma}}} \exp \left(-\frac{\sigma^2 +{M_\sigma}^2}{2 C_{\sigma\sigma}} \right) \cosh \left(\frac{M_\sigma \sigma}{C_{\sigma\sigma}} \right) \ ,
\end{align}
so that we can rewrite Eq. (\ref{eq smallscalegg}) as an integral over a non-negative range,
\begin{align}
\label{eq smallscaleggpm}
\mP &= \int^{+ \infty}_0 d \sigma \ \frac{\mW^\pm}{\sqrt{2 \pi \sigma^2}} \exp \left(-\frac{\vpa^2}{2 \sigma^2} \right) \ .
\end{align}
In this small-scale scenario, at any given position in the universe, the pairs contributing to the corresponding local velocity PDF all belong to the same halo.
We can therefore infer the variance $\sigma^2$, which is the only remaining local parameter, from consideration on the physical properties of a single isolated halo.
A useful discussion about this topic can be found in \citet{sheth1996}, in which, under the assumption that halos are virialized and isothermal systems, an expression for $\sigma=\sigma(M)$ is derived, where $M$ is the mass of the halo.
If we compare Eq. (\ref{eq smallscaleggpm}) with the corresponding expression for the small-scale velocity PDF derived by Sheth, his equation (5), we realize that $\mW^\pm$ is essentially the probability $\mT(M)$ that a pair with separation $r$ belongs to an halo of mass $M$.
More specifically, for any fixed (small) separation $r$, \citet{sheth1996} comes to the following integral,
\begin{equation}\label{eq Sheth}
\mP = \int_0^{+\infty} dM \ \frac{\theta(M-M_{min}) \ \mT(M)}{\sqrt{2\pi {\sigma(M)}^2}} \ \exp \left(-\frac{{v_\parallel}^2}{2{\sigma(M)}^2}\right)
\end{equation}
where $\theta$ is a step function and $M_{min}$ is the minimum halo mass compatible with the separation $r$.
From Eqs. (\ref{eq smallscaleggpm}) and (\ref{eq Sheth}) it follows that $\mW^\pm$ can be seen as a two-parameter ansatz for the function $|dM/d\sigma| \ \theta[M(\sigma) -M_{min}] \ \mT[M(\sigma)]$.
As shown by \citet{sheth1996}, under reasonable physical assumptions, this latter can be computed once a mass function is provided.

A straightforward procedure to include this theoretical prediction for the small-scale limit in our model can be obtained as follows.
From Tab. \ref{tab mvsM} it is easy to see that in the small-scale limit $\mP$ can be expressed as a function of its first two even central moments $c^{(2)}$ and $c^{(4)}$,  
\begin{align}
{M_\sigma}^2 &= \sqrt{\frac{3}{2}{c^{(2)}}^2 - \frac{1}{6}c^{(4)}} \label{eq MsgSSL} \\
C_{\sigma\sigma} &= c^{(2)} -  {M_\sigma}^2 \label{eq CsgsgSSL} \ .
\end{align}
We then need a theoretical prediction of these two moments.
Form Eq. (\ref{eq Sheth}), it follows 
\begin{equation}
c^{(n)} = \int_{M_{min}}^{+\infty} dM \ 3^ \frac{n-2}{2} {\sigma(M)}^n \ \mT(M)
\end{equation}
for $n=2,4$, which completes the modelling. 

It should be noted that if we want to adapt the model introduced in section \ref{sec strategy} to the small-scale limit just discussed, a decreasing (or even flat, as proposed in Sec. \ref{sec simple}) profile for the function $\varphi_{C\mu\mu}$ is no longer acceptable.
More explicitly, a decreasing profile implies that if $\lim_{r \rightarrow 0} C_{\mu\mu}=0$, then $C_{\mu\mu}=0$ at any separation, which means $C_{\mu\sigma}=0$ at any separation as well.
A more general profile for the functions $\varphi$ is then required.
The simplest possible improvement is to define a scale $r_3$ below which $\varphi_{C\mu\mu}$, more precisely its parallel and perpendicular components, is damped.
Since, based on the discussion in Sec. \ref{sec skew problem}, we
expect this scale to roughly correspond the skewness maximum, from the
right panels of Fig. \ref{fig moms multi} we can argue that $r_3 \sim 5h^{-1}$Mpc.

\section{Precision of the model}\label{app sys error}

\begin{figure*}
 \begin{center}
   \includegraphics[width=18cm]{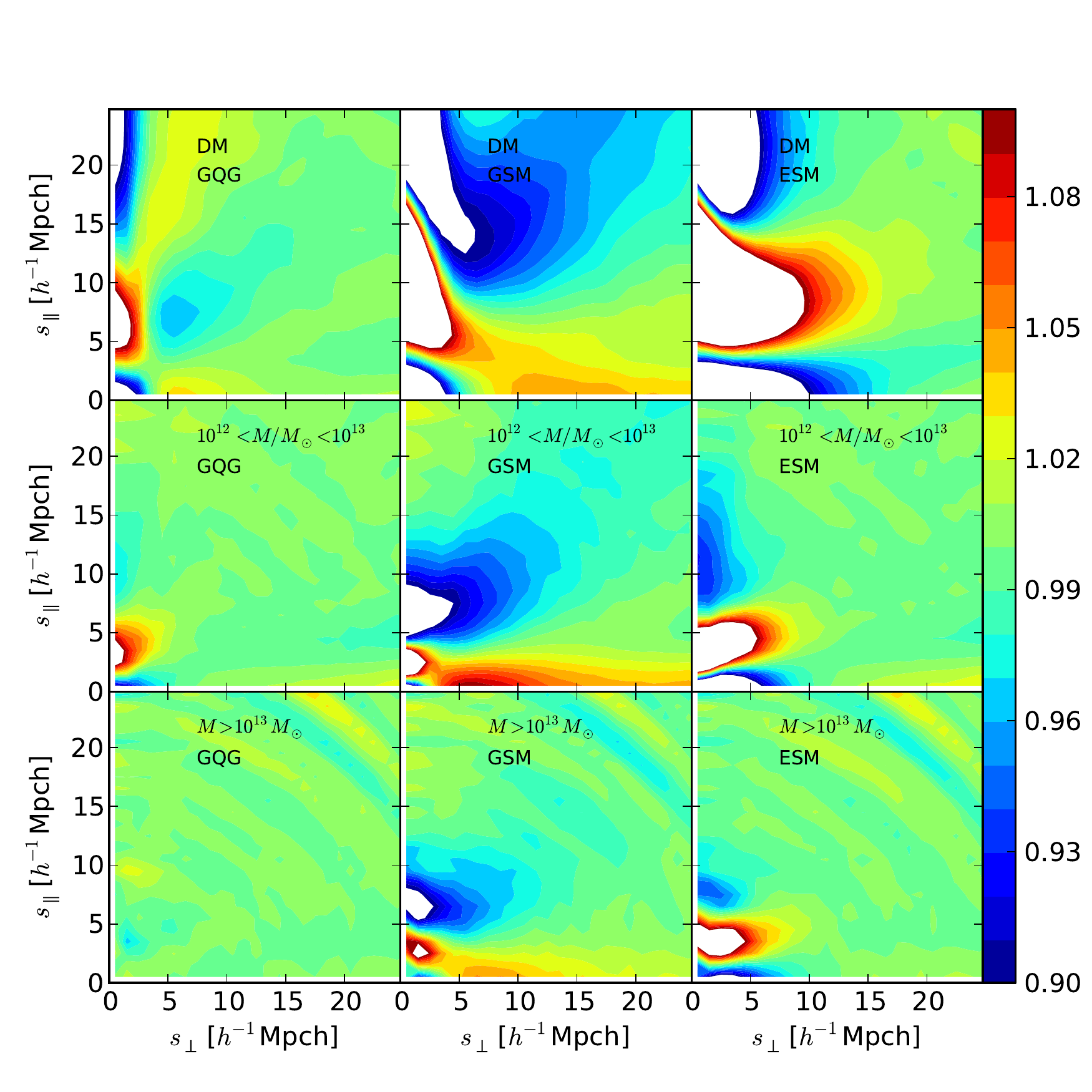}
    \caption{Colour map of the ratio, at $z=0$, between the two-dimensional correlation function obtained via GQG (left panels), GSM (central panels) and ESM (right panels) with respect to that obtained by measuring the velocity PDF directly from the simulations, for different tracers as labeled in the figure.}
  \label{fig xis_ratio_multi}
 \end{center}
\end{figure*}

For completeness, in Fig. \ref{fig xis_ratio_multi} we explicitly show the ratio between the $\xi_S(s_\perp, s_\parallel)$ corresponding to the three models discussed in this work, GQG, GSM and ESM (red, green and blue solid lines in Fig. \ref{fig xis_multi}), with respect to the reference one obtained by measuring the velocity PDF directly from the simulations (black dashed lines in Fig. \ref{fig xis_multi}).
Only $z=0$ is considered, but different redshifts yield similar results.
As expected, GQG outperforms GSM and ESM, being percent accurate almost everywhere.
The residual small-scale discrepancies can in principle be removed by improving the modelling of the $\varphi$ functions.
We leave this topic to further work.

\bsp	
\label{lastpage}
\end{document}